\documentclass[journal]{IEEEtran}
\ifCLASSINFOpdf
\else
\fi
\usepackage{amsmath,amssymb,amsfonts}
\usepackage{algorithmic}
\usepackage{textcomp}
\usepackage{comment}

\usepackage{stackengine}
\usepackage{graphicx}      
\usepackage{multirow}
\usepackage{float}
\usepackage{varwidth}
\usepackage{amsmath}
\usepackage{amssymb}
\usepackage{graphicx, epstopdf}
\usepackage{multicol}
\usepackage{array}
\usepackage{csquotes}
\newcolumntype{P}[1]{>{\centering\arraybackslash}p{#1}}
\usepackage{breqn}
\usepackage{setspace}
\usepackage{verbatim}
\usepackage{url}
\usepackage[dvipsnames]{xcolor}

\newtheorem{lemma}{Lemma}
\newtheorem{theorem}{Theorem}

\newtheorem{definition}{Definition}
\usepackage{cleveref}
\usepackage{subfigure}
\usepackage{wrapfig}
\usepackage[rflt]{floatflt}

\begin{document}
%
\title{Region of Attraction Estimation Using Union Theorem in Sum-of-Squares Optimization}
\author{Bhaskar Biswas, Dmitry Ignatyev, Argyrios Zolotas and Antonios Tsourdos	
\thanks{Bhaskar Biswas, Ph.D. Student, School of Aerospace, Transport and Manufacturing, Cranfield University, U.K. (e-mail: bhaskar.biswas@cranfield.ac.uk)}
\thanks{Dmitry Ignatyev, Senior Research Fellow, School of Aerospace, Transport and Manufacturing, Cranfield University, U.K. (e-mail: D.Ignatyev@cranfield.ac.uk)}
\thanks{Argyrios Zolotas, Reader, School of Aerospace, Transport and Manufacturing, Cranfield University, U.K. (e-mail: A.Zolotas@cranfield.ac.uk)}
\thanks{Antonios Tsourdos, Professor, School of Aerospace, Transport and Manufacturing, Cranfield University, U.K. (e-mail: a.tsourdos@cranfield.ac.uk)}
}

\markboth{Preprint, May~2023}%
{Biswas \MakeLowercase{\textit{et al.}}: RoA Estimation Using Union
Theorem in SOS Optimization}
%



\maketitle

\begin{abstract}
 Appropriate estimation of Region of Attraction for a nonlinear dynamical system plays a key role in system analysis and control design. Sum-of-Squares optimization is a powerful tool enabling Region of Attraction estimation for polynomial dynamical systems. Employment of a positive definite function called shape function within the Sum-of-Squares procedure helps to find a richer representation of the Lyapunov function and a larger corresponding Region of Attraction estimation. However, existing Sum-of-Squares optimization techniques demonstrate very conservative results. The main novelty of this paper is the Union theorem which enables the use of multiple shape functions to create a polynomial Lyapunov function encompassing all the areas generated by the shape functions. The main contribution of this paper is a novel computationally-efficient numerical method for Region of Attraction estimation, which remarkably improves estimation performance and overcomes limitations of existing methods, while maintaining the resultant Lyapunov function polynomial, thus facilitating control system design and construction of control Lyapunov function with enhanced Region of Attraction using conventional Sum-of-Squares tools. A mathematical proof of the Union theorem along with its application to the numerical algorithm of Region of Attraction estimation is provided. The method yields significantly enlarged Region of Attraction estimations even for systems with non-symmetric or unbounded Region of Attraction, which is demonstrated via simulations of several benchmark examples.   
\end{abstract}
\section{Introduction}
\label{sec:introduction}
\IEEEPARstart{T}{he} stability analysis and design of target-oriented controllers or policies for autonomous systems are of significant importance in various fields such as aerospace, autonomous driving, and robotics. One popular approach for stabilizing a system is linearizing a nonlinear system and eliminating nonlinearities. However, utilizing the unique characteristics of a nonlinear system can lead to better performance. Many nonlinear systems are only stable in a specific region around an equilibrium point (EP). This region, known as the Region of Attraction (ROA) of the relevant EP, is crucial for system stability, and robustness and determines the extent to which initial states can be disturbed from the expected steady states, \cite{r1}. Knowing the stable region is essential for system analysis, design and clearance of control systems. Determining the exact ROA for a nonlinear system can be a challenging task, made more difficult by the use of nonlinear filters, adaptive control, or parameter estimation strategies. There are several promising novel approaches including Lyapunov function (LF) based ones that provide a route for stability analysis and control design.\\[3pt]
The Lyapunov stability theory forms the basis of the LF-based method. Although the process of creating LF has been extensively researched \cite{r2}, it can be a challenging task and may not be applicable to all cases. To create a more comprehensive approach, numerous computational construction methods have been developed. These methods employ various forms of relaxation to develop a generalized approach for LF construction. One such approach is the sum-of-squares (SOS) optimization technique, which can simultaneously estimate the ROA and construct an LF for polynomial dynamical systems.  In this paper, we propose a method using the SOS optimization technique to get a better estimation of the ROA for polynomial dynamical systems.\\[3pt]
An SOS polynomial is a non-negative polynomial, which can be expressed as a sum of squares of other polynomials. An LF can be represented as an SOS polynomial as both are non-negative. The SOS approaches generalizes linear matrix inequality (LMI). A detailed discussion about LMI and its application to estimate the ROA can be found in \cite{r4} and \cite{r5}. In \cite{r6}, the authors presented an algorithm to check if a real polynomial is an SOS polynomial and find its representation if so. SOS is a relaxation procedure, and the conditions required to express a non-negative polynomial as a relaxed SOS polynomial have been discussed in \cite{r7} and \cite{r8}. Tutorials on SOS optimization technique, including MATLAB code for estimating ROA, have been presented in \cite{r9}, \cite{r10} and \cite{r11}. The algorithm in \cite{r12} can certify the non-existence of a common quadratic LF for a switched system. In \cite{r32}, the authors have presented an algorithm to solve the nonconvex SOS problems efficiently. Tools such as SOSOPT \cite{r25}, BiSOS \cite{r32} and SOSTOOLS \cite{r31} can be used to solve SOS optimization-related problems. Further information on the SOS optimization method can be found in \cite{r13}, \cite{r14}, and \cite{r15}.\\[3pt]
The works of \cite{r17}, \cite{r18}, \cite{r19} and \cite{r20} illustrate the application of the SOS optimization technique to design the controller alongside the estimation of ROA. In \cite{r21}, \cite{r22}, and \cite{r34}, the authors estimated the ROA for the aircraft attractors using the SOS optimization method.  These authors employed a method based on Shape Function (SF) to obtain a larger ROA. This method estimates a very large ROA. However, the method gives conservative result if the actual ROA is non-symmetric. To overcome this problem, in \cite{r23}, authors proposed a technique called RcomSSF that utilizes SOS optimization to estimate a larger ROA for all the systems including non-symmetric ROA. However, the resulting LF from this method is not a polynomial, limiting its application in further studies using SOS optimization techniques such as control system design and construction of control Lyapunov function (CLF).\\[3pt]
This paper bypasses the above limitations and proposes a novel SOS-based method that significantly increases ROA estimation while maintaining the resultant LF polynomial. The key enabler is the proposed Union theorem which enables the use of multiple SFs to create a polynomial LF encompassing all the areas generated by the SFs. The paper provides the mathematical proof of the Union theorem and demonstrates its application to the numerical algorithm of ROA estimation.
The effectiveness of the proposed method is demonstrated using four examples, representing both 2-D and 3-D systems, with bounded or unbounded, symmetric or non-symmetric ROAs.\\[3pt]
The paper is structured as follows: in Section 2, the background information on ROA and SOS optimization is provided, and the problem is formulated. In Section 3, the Union theorem and its application to estimate the ROA are discussed. Section 4 covers the numerical process used to solve the proposed method. Section 5 explains the selection of the required SF. The simulation results are presented in Section 6. Finally, in Section 7, the paper concludes with final thoughts.
\section{Preliminaries and Problem Formulation}
\subsection{Region of Attraction}
\begin{figure}[H]
		\begin{center}
			\includegraphics[scale=.55]{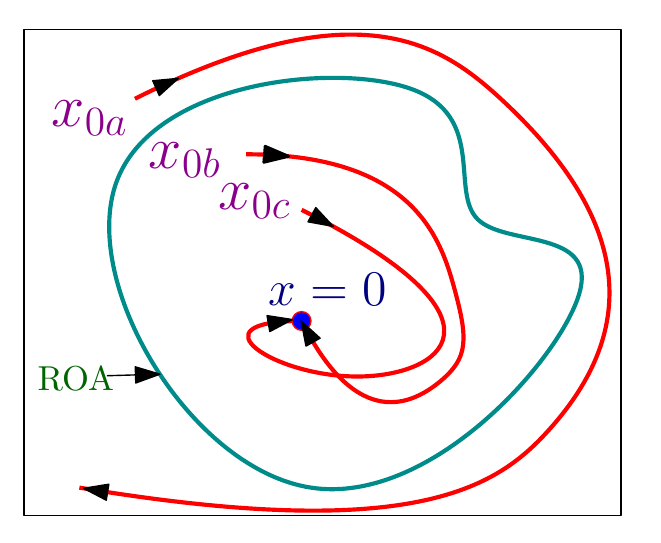}
			\caption{Region of Attraction}
			\label{1ms}
		\end{center}
	\end{figure}

ROA is a domain surrounding the EP, which includes the EP. When the trajectory starts within this domain, it will eventually reach the EP as time progress. It has been shown in Fig.~\ref{1ms}. Let us consider the following autonomous nonlinear system,
\begin{align*}
	\dot{x}=f(x), \hspace{10pt} x(0)=x_0,\hspace{4pt}\text{where,}
\end{align*}
	\begin{itemize}
		\item{$D\subset{\mathbb{R}^n}$: Domain contains origin}
		\item{$f(x):D\rightarrow{\mathbb{R}^n}$: Locally Lipschitz}
		\item{$\phi(x)$: Solution of the system}
	\end{itemize}
	Without loss of generality, we can assume that $x=0$ is the origin of the system and asymptotically stable.
\begin{definition}[ROA]The ROA of origin, $x=0$ is,
	\begin{align*}
		R_A=\{x\in{D}:\phi(x)\hspace{3pt}\text{is}\hspace{3pt} \forall\hspace{3pt}{t\geq{0}}\hspace{3pt} \text{and} \lim_{t\to\infty} \phi(x)\rightarrow{0}\}\\[-20pt]
	\end{align*}	
	If $x=0$ is an asymptotically stable EP for the system, then its ROA, $R_A$ is an open, connected, and invariant set. Moreover, the boundary of $R_A$ is formed by trajectories, \cite{r1}.\vspace{2pt}
\end{definition}
\begin{lemma}
	If there exist a continuously scalar function $V(x):\mathbb{R}^n\rightarrow\mathbb{R}$ and a positive scalar $\gamma\in\mathbb{R}^+$, such that,
	\begin{subequations}
		\begin{align}
			&V(x)>0 \hspace{4pt}\forall\hspace{3pt} x\neq 0\hspace{2pt} \text{and} \hspace{2pt}V(0)=0\label{1a}\\[2pt]
			&\Omega_\gamma:=\{x:V(x)\le\gamma\}\hspace{3pt}\text{is bounded}\label{1b}\\[2pt]
			&\Omega_\gamma\subseteq\{x:\frac{\partial V}{\partial x}f(x)<0\}\cup \{0\}\label{1c}
		\end{align}
	\end{subequations}
	Then the origin is asymptotically stable and $\Omega_\gamma$ is a subset of ROA \cite{r33}.\vspace{2pt}	
\end{lemma}
The system is globally asymptotically stable if $\gamma$ is unbounded. The above lemma provides a strict LF, $V(x)$, by bounding the value of $\gamma$. Optimizing $\gamma$ gives the maximum estimate of $\Omega_\gamma$ and the SOS optimization method is one of the ways to do it. The method simultaneously yields a SOS function $V(x)$ and optimized value of $\gamma$.
\subsection{Sum-of-Squares Optimization}
SOS polynomials are non-negative polynomials and can be expressed as a sum of squares of other polynomials. SOS optimization is a useful technique for getting a LF as both the SOS polynomial and LF  are non-negative. Familiarity with definitions and lemmas in references \cite{r17} and \cite{r18} is necessary for converting \emph{Lemma 1} into an SOS optimization problem.\vspace{2pt}
\begin{definition}[Monomial]
	Monomial is a function of $n$ variables defined as $m_\alpha(x)=x^\alpha:=x_1^{\alpha_1}x_2^{\alpha_2}...\hspace{2pt} x_n^{\alpha_n}\hspace{2pt} \forall\hspace{2pt} \alpha\in \mathbb{Z}_+^n$. The degree of a monomial is defined, deg  $m_\alpha:=\sum_{i=1}^{n}\alpha_i$.
\end{definition}\vspace{2pt}
\begin{definition}[Polynomial] 
A $n$ variables polynomial can be expressed as finite linear combination of monomials,
 $f:=\displaystyle\sum_\alpha c_\alpha m_\alpha=\displaystyle\sum_\alpha c_\alpha x^\alpha \hspace{3pt}\forall\hspace{3pt} c_\alpha\in \mathbb{R}$.
	The set of all polynomials in $n$ variables is defined as $ \mathcal{R}_n$. The degree of $f$ is defined as deg $f:=max_\alpha$ deg $m_\alpha$ with $c_\alpha\not=0$.
\end{definition}\vspace{2pt}
\begin{definition}[Positive Semi-definite Polynomial]
	A polynomial, $f(x)$, is called Positive Semi-definite Polynomial (PSD) if $f(x)\geq0\hspace{5pt} \text{for all}\hspace{5pt} x\in\mathbb{R}^n$.
\end{definition}\vspace{2pt}
\begin{definition}[Sum-of-Squares (SOS) polynomial]
	A polynomial, $s(x)\in\mathcal{R}_n$ is a SOS polynomial if there exist polynomials $f(x)_i\in\mathcal{R}_n$ such that\vspace{-5pt}
	\begin{align*}
		s(x)=\displaystyle\sum_{i=1}^t f_i^2,\hspace{2pt} f_i\in \mathcal{R}_n, i=1, ...,t
	\end{align*}
	Again, s(x) is a SOS if there exists $Q\geq0$ such that $s(x)=Z^TQZ$, where Z is the vector of monomials.
	The set of SOS polynomials in n variables is defined as
	\begin{align*}
		\mathcal{\sum}_n:=\bigg\{s(x)\in  \mathcal{R}_n\hspace{2pt}\bigg |\hspace{2pt} s(x)=\displaystyle\sum_{i=1}^t f_i^2,\hspace{2pt} f_i\in \mathcal{R}_n, i=1, ...,t\bigg\}
	\end{align*}
	Now, always $s(x)\geq 0$ if $s(x)\in\sum_n\forall\hspace{2pt} x\in \mathbb{R}^n.$
\end{definition}\vspace{2pt}
\begin{definition}[Multiplicative Monoid]
	Let us consider that given $\{g_1,...g_t\}\in \mathcal{R}_n.$ The set of all finite products of $g_j$'s including 1 (i.e. the empty product) is known as Multiplicative Monoid. It is denoted as $\mathcal{M}(g_1,...,g_t)$. For completeness define $\mathcal{M}(\phi):=1.$
	For example, $\mathcal{M}(g_1,...g_t)=\Big\{g_1^{k_1}g_2^{k_2}\hspace{2pt}\big |\hspace{2pt} k_1, k_2\in\mathbb{Z}_+\Big\}$
\end{definition}\vspace{2pt}
\begin{definition}[Cone]
	Let us consider that given $\{f_1,$ $...,f_r\}\in \mathcal{R}_n.$ The cone based on $f_i$'s is defined by,
	\begin{align*}
		&\mathcal{P}(f_1,...,f_r):=\bigg\{s_0+\displaystyle\sum_{i=1}^l s_{i}b_i\hspace{2pt}\bigg |\hspace{2pt} l\in \mathbb{Z}_+,s_i\in\mathcal{\sum}_n,\\
		&\hspace{140pt}b_i\in \mathcal{M}(f_1,...,f_r)\bigg\}
	\end{align*}
	It is noted that if $s\in\mathcal{\sum}_n$ and $f\in\mathcal{R}_n$ then $f^2s\in\mathcal{\sum}_n.$ So, a cone of  $\{f_1,...,f_r\}$ can be expressed as a sum of $2^r$ terms. For example, $\mathcal{P}(f_1,f_2)=\{s_0+s_1f_1+s_2f_2+s_3f_1f_2\hspace{2pt}|\hspace{2pt}s_0,...,s_3\in\mathcal{\sum}_n\}$
\end{definition}\vspace{2pt}
\begin{definition}[Ideal]
	Let us consider that given $\{h_1,$ $...,h_u\}
	\in \mathcal{R}_n.$ The Ideal based on $h_k$'s is defined by,
	\begin{align*}
		\mathcal{I}(h_1,...,h_u):=\left\{\displaystyle\sum^u_{k=1} h_{k}f_k\hspace{2pt}\Big |\hspace{2pt} f_k\in\mathcal{R}_n\right\}
	\end{align*}
\end{definition}\vspace{2pt}
Now, the Positivstellensatz theorem based on the above definitions is stated below,\vspace{2pt}
\begin{theorem}[Positivstellensatz (P-staz)]
	Let us consider the given polynomials $\{f_1,...,f_r\}$, $\{g_1,...,g_t\}$ and $\{h_1,...,h_u\}$ are in $\mathcal{R}_n.$ Then the below statements are equivalent,\\[5pt]
	\begin{subequations}
1. The set
		\begin{align}
			X=\left\{x\in\mathbb{R}^n\left|
			\begin{array}{@{}ll@{}}
				f_1(x)\geq0,...,f_r(x)\geq0\\
				g_1(x)\not=0,...,g_t(x)\not=0\\
				h_1(x)=0,...,h_u(x)=0
			\end{array}\right.\right\}\text{is empty.}\label{2}\vspace{2pt}
		\end{align}
2. There exist polynomials $f_p\in\mathcal{P}\{f_1,...,f_r\},$\hspace{3pt}$g_p\in\mathcal{M}\{g_1,$ $...,g_t\}$
		and $h_p\in\mathcal{I}\{h_1,...,h_u\}$ such that
		\begin{align}
			f_p+g_p^2+h_p=0\label{2b}
		\end{align}
	\end{subequations} 
\end{theorem}\vspace{2pt}
The P-staz guarantees the existence of infeasibility certificates, given by polynomials $f$, $g$ and $h$.
It can be used to convert the ROA estimation problem (\emph{Lemma 1}) into an SOS optimization problem, but it's complex to solve. To simplify it, we need to transform it into a relaxation form called S-procedure, which is part of set containment. In the next section, we'll give a quick explanation of set containment.
\subsection{Set Containment}
\begin{figure}[H]
		\begin{center}
			\includegraphics[scale=.25]{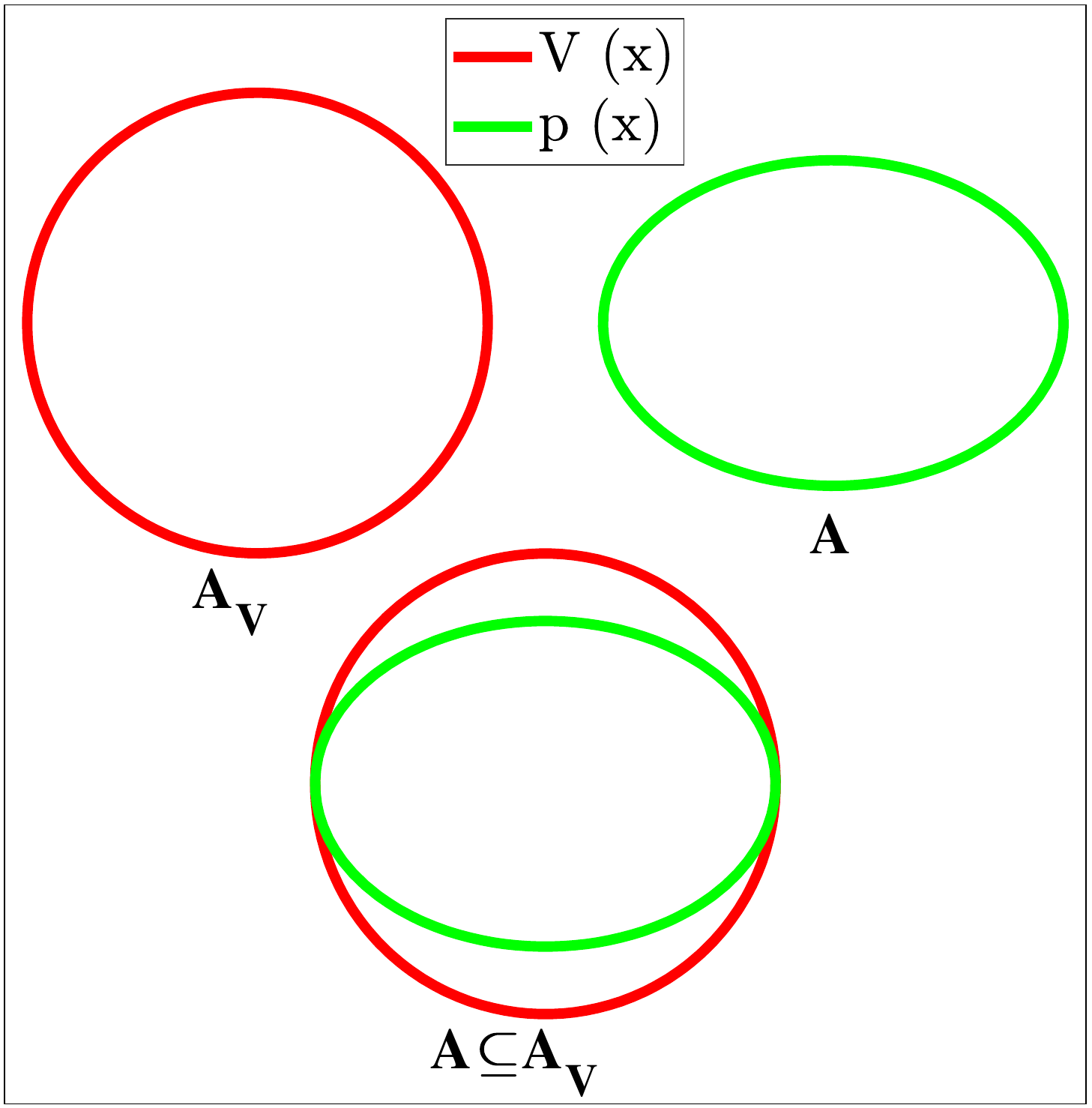}
			\caption{Set Containment}
			\label{2ms}
		\end{center}
	\end{figure}

If a particular set is entirely inside or a subset of another set then it is referred to as the latter set containing the former set. Set containment plays a crucial role in nonlinear control theory as various nonlinear analysis problems can be expressed as set containment constraints. Consider the polynomials $V(x)$ and $p(x)$ that define the following sets:
\begin{subequations}
	\begin{align}
		&A_V:= \{x\in \mathbb{R}^n: V(x)\leq0\}\label{3a}\\[2pt]	
		&A:= \{x\in \mathbb{R}^n: p(x)\leq0\}\label{3b}
	\end{align}
\end{subequations}
	The sets $A_V$ and $A$ are shown in Fig.~\ref{2ms} by red and green colour, respectively. If we want that $A$ should be always inside the $A_V$ ($A \subseteq A_V$) then some conditions are required to satisfy by $A$ and $A_V$. The S-procedure gives these conditions, \cite{r9}.
\subsection{S-Procedure}
The S-procedure is a relaxation method that uses a specific condition to create a quadratic inequality from other quadratic inequalities. It turns a non-LMI system of quadratic inequalities into an LMI, \cite{r28}.  It is commonly used in control theory to prove set containment involving quadratic or higher-degree polynomials.\vspace{2pt}
\begin{theorem}[S-Procedure]
	Suppose that $V(x)$ and $p(x)$ are quadratic functions and its enclosed regions are defined by \eqref{3a} and \eqref{3b}, respectively. Again consider that there exist matrices $B_0, B_1\in\mathbb{R}^{n+1\times n+1}$ such that
	\begin{align*}
		V(x) = \left[ {\begin{array}{*{20}{c}}
				{1}\\
				{x}
		\end{array}} \right]^T B_0 \left[ {\begin{array}{*{20}{c}}
				{1}\\
				{x}
		\end{array}} \right],\hspace{5pt}
		p(x) = \left[ {\begin{array}{*{20}{c}}
				{1}\\
				{x}
		\end{array}} \right]^T B_1 \left[ {\begin{array}{*{20}{c}}
				{1}\\
				{x}
		\end{array}} \right]
	\end{align*}
	If there exist scalar $s\geq0$ such that $-B_0+sB_1\succeq0$, then $A \subseteq A_V$, \cite{r9}. 
\end{theorem}\vspace{2pt}
\begin{theorem}[Generalised S-Procedure]
	Let us consider that the given polynomials $\{V(x), p(x)_1, p(x)_2,...,$ $p(x)_m\}\in\mathcal{R}_n$. Now if there exist polynomials $\{s_k\}^m_{k=1}\in\mathcal{\sum}_n$ such that $V-\sum_{k=1}^{m}s_kp_k=q$ with $q\in\mathcal{\sum}_n$, then, (\cite{r29}),
	\begin{align*}
		\Bigg\{\bigcap_{k=1}^m\{x\in\mathbb{R}^n\hspace{2pt}|\hspace{2pt}p_k(x)\geq0\}\Bigg\}\subseteq\{x\in\mathbb{R}^n\hspace{2pt}|\hspace{2pt}V(x)\geq0\}
	\end{align*}
\end{theorem}\vspace{4pt}
The Theorem tells us that the intersection region defined by $\Big\{\bigcap_{k=1}^m\{x\in\mathbb{R}^n\hspace{2pt}|\hspace{2pt}p_k(x)\geq0\}\Big\}$ will be inside or a subset of the region defined by $\{x\in\mathbb{R}^n\hspace{2pt}|\hspace{2pt}V(x)\geq0\}$.
The Theorem from S-Procedure can be written in the P-staz form. If we express the \emph{Theorem 3} in P-staz form, then it is required to check whether or not the below set is empty, (\cite{r17}).
\begin{align*}
	A_{VP}=\left\{x\in\mathbb{R}_n\hspace{2pt}\Bigg|\hspace{2pt}
	\begin{array}{@{}ll@{}}
		-V(x)\geq0,p(x)_1\geq0,...,p(x)_m\geq0\\[2pt]
		V(x)\not=0
	\end{array}\right\}
\end{align*}
Using point 2 of \emph{Theorem 1}, we can define a condition when the above set is empty. Let us define the below cone and multiplicative monoid,
\begin{subequations}
	\begin{align}
		&f_p=q(-V)+\sum_{k=1}^{m}s_k(-V)p_k\in\mathcal{P}(-V,p_1,...,p_m)\label{4a}\\[2pt]
		&g_p=V\in\mathcal{M}(V)\label{4b}\\[2pt]
         &\text{Verifying $f_p+g_p^2=0$ (point 2 of \emph{Theorem 1}),}\nonumber\\[2pt]
       &f_p+g_p^2=q(-V)+\sum_{k=1}^{m}s_k(-V)p_k+V^2\nonumber
	\end{align}
 \end{subequations} 
\begin{align}
		&\text{[Using the value of $f_p$ and $g_p$ from \eqref{4a} and \eqref{4b}]}\nonumber\\[2pt]
		&=-(V-\sum_{k=1}^{m}s_kp_k)V-\sum_{k=1}^{m}s_kVp_k+V^2=0\nonumber
	\end{align}
The set $A_{VP}$ is proven to be empty by $f_p$ and $g_p$. This proves that the S-Procedure can be written in P-staz form. Again the S-Procedure is a simpler form of P-staz and can be obtained from it. The S-Procedure is commonly used to verify the positive semi-definiteness of polynomial systems as it simplifies the computation process by converting the system into LMI.
\subsection{SOS Optimization Form of ROA}
To get a better estimation of ROA, we need to optimize $\gamma$ of \emph{Lemma 1}. Using \emph{Theorem 1}, \emph{Lemma 1} can be written in the SOS optimization form, \cite{r17}. First, we can write \emph{Lemma 1} in the below optimization form,\vspace{0.1pt}
\begin{align}
	&\gamma^*=\max\gamma\nonumber\\[2pt]
	&
	\begin{array}{@{}ll@{}}
		\text{Subject to}:\{x\in\mathbb{R}^n:V(x)>0, \hspace{2pt} x\neq 0\}\hspace{2pt} \text{and} \hspace{2pt}V(0)=0\vspace{2pt}\\[2pt]
		\{x\in\mathbb{R}^n:V(x)\le\gamma,\hspace{2pt}x\neq 0\}\subseteq
		\{x\in\mathbb{R}^n:\frac{\partial V}{\partial x}f(x)<0\}
	\end{array}\label{5}
\end{align}
Eq. \eqref{5} can be converted into the following form which is amenable to the P-staz form-\eqref{2}.
\begin{subequations}
	\begin{align}
		&\gamma^*=\max\gamma\nonumber\\[2pt]	
		&\text{Subject to}:
		\left\{x\in\mathbb{R}^n\bigg|\hspace{2pt}
		\begin{array}{@{}ll@{}}
			-V(x)\geq0\\
			l_1(x)\neq 0
		\end{array}\right\}\hspace{3pt}\text{is empty}\label{6a}\\[2pt]
		&\left\{x\in\mathbb{R}^n\bigg|\hspace{2pt}
		\begin{array}{@{}ll@{}}
			\gamma-V(x)\geq0,\hspace{2pt}\frac{\partial V}{\partial x}f(x)\geq0\\[2pt]
			l_2(x)\neq 0
		\end{array}\right\}\hspace{3pt}\text{is empty}\label{6b}
	\end{align}\label{6}
\end{subequations}
To satisfy $V(0)=0$, the constant term is set to zero. The non-polynomial constraint $x\neq0$ in equation \eqref{5} is replaced with two fixed, positive definite SOS polynomials $l_1$ and $l_2$ as P-staz works only with polynomials. To apply point 2 of \emph{Theorem 1}, a cone and multiplicative monoid need to be created.  The cone and multiplicative monoid of equation \eqref{6a} are,
\begin{subequations}
	\begin{align}
		&f_{p1}=s_1+(-V(x))s_2\in\mathcal{P}(-V)\\[2pt]
		&g_{p1}=l_1^{k_1}\in\mathcal{M}(l_1);
	\end{align}\label{7}
\end{subequations}
The Cone and Multiplicative Monoid of \eqref{6b} are,
\begin{subequations}
	\begin{align}
		&f_{p2}=s_3+(\gamma-V(x))s_4+\frac{\partial V}{\partial x}f(x)s_5\nonumber\\[2pt]
		&\hspace{10pt}+(\gamma-V(x))\frac{\partial V}{\partial x}f(x)s_6\in\mathcal{P}\bigg((\gamma-V),\hspace{2pt}\frac{\partial V}{\partial x}f\bigg)\\[2pt]
		&g_{p2}=l_2^{k_2}\in\mathcal{M}(l_2)
	\end{align}\label{8}\vspace{-1pt}
	Substituting \eqref{7} and \eqref{8} in \eqref{2b} of P-staz, we get,
\end{subequations}
\begin{subequations}
	\begin{align}
		s_1+(-V(x))s_2+l_1^{2k_1}=0\label{9a}
	\end{align}	
	\begin{align}
		&s_3+(\gamma-V(x))s_4+\frac{\partial V}{\partial x}f(x)s_5
		+(\gamma-V(x))\frac{\partial V}{\partial x}f(x)s_6\nonumber\\[2pt]
		&\hspace{150pt}+l_2^{2k_2}=0\label{9b}
	\end{align}\label{9}
\end{subequations}
Eq. \eqref{9} is too complex for direct solving by SOS programming tools. Hence, we need to modify it for numerical algorithms. We treat $s_i$ and $k_i$ as decision variables, some of which can be assigned convenient values. To avoid infeasibility, we set $k_1=k_2=1$, the next smallest value after zero. Also, we choose $s_1=\hat{s}_1l_1, s_2=l_1, s_3=\hat{s}_3l_2, s_4=s_0l_2, s_5=l_2$ and $s_6=0$. Here, $\hat{s}_1, \hat{s}_3$, and $s_0$ are SOS polynomials. By substituting all the required values of $s_i$ and $k_j$ into Eq. \eqref{9a}, we get,
\begin{subequations}
	\begin{align}
		&\hat{s}_1l_1-V(x)l_1+l_1^{2\times1}=0\nonumber\\[2pt]
		&V(x)-l_1=\hat{s}_1\nonumber\\[2pt]
		&V(x)-l_1\in\mathcal{\sum}_n \hspace{8pt}[\text{as}\hspace{4pt} \hat{s}_1\hspace{3pt}\text{is a SOS polynomial}]\label{10a}
	\end{align}
	Similarly, after simplification of \eqref{9b}, we get,
	\begin{align}
		-\bigg[\frac{\partial V}{\partial x}f(x)+l_2\bigg]+(V(x)-\gamma)s_0\in\mathcal{\sum}_n\label{10b}
	\end{align}\label{10}
\end{subequations}
The Eq. \eqref{9} has been converted from P-staz into S-procedure in Eq. \eqref{10}. The SOS optimization form of the problem in Eq. \eqref{5} can be written as,
\begin{subequations}
	\begin{align}
		&\gamma^*=\max_{\mathbf{s_0\in\mathcal\sum_n}}\gamma\nonumber\\[2pt]	
		&\text{Subject to}: \hspace{2pt}V(x)-l_1\in\mathcal{\sum}_n\label{11a}\\[2pt]
		-&\bigg[\frac{\partial V}{\partial x}f(x)+l_2\bigg]+(V(x)-\gamma)s_0\in\mathcal{\sum}_n\label{11b}
	\end{align}\label{11}
\end{subequations}
The above equation estimates a very small ROA, as discussed and demonstrated in the simulation section using the Van der Pol system. To improve the ROA estimation, a new function called shape function has been introduced in \cite{r17} along with Eq. \eqref{11}, and we will discuss this further in the next section.
\subsection{Shape Function}
Shape function (SF) is a function whose encircled region is surrounded by the bounded LF, and it forces the bounded LF to increase its size by increasing its own (SF) size in every iteration until any part of the LF touches the unstable region. In simpler terms, the SF determines the final shape and size of the bounded LF. The introduction of the SF in P-staz or SOS optimization method can be done in the following way, \cite{r17},
\begin{definition}[Shape Function]
	SF is a  positive definite function. The variable size region under it, is defined as,
\end{definition}
\begin{align}
	\varepsilon_\beta=\{x:p(x)\le\beta\}\subseteq\Omega_\gamma\label{12}		
\end{align}
Using \eqref{1b} in \eqref{12}, we can write,
\begin{align}
	&\varepsilon_\beta=\{x\in\mathbb{R}^n:p(x)\le\beta\}\subseteq\{x\in\mathbb{R}^n:V(x)\le\gamma\}\label{13}	
\end{align}
Again, the optimization form of \eqref{13} is,
\begin{align}
	&\beta^*=\max\beta\nonumber\\[2pt]	
	&\text{Subject to}:\nonumber\\[2pt]
	&\{x\in\mathbb{R}^n:p(x)\le\beta\}\subseteq\{x\in\mathbb{R}^n:V(x)\le\gamma\}\label{14}
\end{align}
Eq. \eqref{14} can be converted into the following form which is amenable to the \eqref{2} of P-staz.\vspace{0.01pt}
\begin{align}
	&\beta^*=\max\beta\nonumber\\[2pt]	
	&\text{Subject to}:\nonumber\\
	&\left\{x\in\mathbb{R}^n\Bigg|\hspace{2pt}
	\begin{array}{@{}ll@{}}
		\beta-p(x)\geq0,\hspace{2pt}-(\gamma-V(x))\geq0\\
		\gamma-V(x)\neq0
	\end{array}\right\}\hspace{3pt}\text{is empty}\label{15}
\end{align}
The cone and multiplicative monoid of \eqref{15} are,
\begin{subequations}
	\begin{align}
		&f_{p3}=s_7+(\beta-p(x))s_8+(-(\gamma-V(x))s_9\nonumber\\[2pt]
		&+(\beta-p(x))(-(\gamma-V(x)))s_{10}\in\mathcal{P}\left((\beta-p),-(\gamma-V)\right)\label{16a}\\
		&g_{p3}=(\gamma-V(x))^{k_3}\in\mathcal{M}\left(\gamma-V\right)\label{16b}
	\end{align}\label{16}
\end{subequations}
Substituting \eqref{16a} and \eqref{16b} in \eqref{2b} of P-staz, we get,\vspace{0.1pt}
\begin{align}
	&s_7+(\beta-p(x))s_8+(-(\gamma-V(x))s_9+\nonumber\\[4pt]
	&(\beta-p(x))(-(\gamma-V(x)))s_{10}+(\gamma-V(x))^{2k_3}=0\label{17}
\end{align}
Eq. \eqref{17} cannot be directly optimized using SOS programming tools due to the presence of $(\gamma-V(x))^{2k_3}$. To make it an optimizable form, we need to choose suitable values for the decision variables $s_i$ and $k_3$. To avoid infeasibility, we set $k_3=1$, the next smallest value after zero. Additionally, we choose $s_7=0$ and $s_8=0$, and by substituting these values and factoring out a $\gamma-V(x)$ term from Eq. \eqref{17}, we get,
\begin{align}
	&0+(\beta-p(x))\times0+(-(\gamma-V(x))s_9\nonumber\\[2pt]
	&+(\beta-p(x))(-(\gamma-V(x)))s_{10}+(\gamma-V(x))^{2\times1}=0\nonumber\\[2pt]
	&-(V(x)-\gamma)+(p(x)-\beta)s_{10}=s_9\nonumber\\[2pt]
	&-(V(x)-\gamma)+(p(x)-\beta)s_{10}\in\mathcal\sum_n\hspace{5pt}\left[\text{as}\hspace{3pt}s_{9}\in\mathcal\sum_n\right]\label{18}
\end{align}
It can be noticed that \eqref{17} transformed from P-staz into S-procedure in  \eqref{18}. The SOS optimization form of the problem \eqref{14}, can be written in the following way,
\begin{align}
	&\beta^*=\max_{\mathbf{s_{10}\in\mathcal\sum_n},\hspace{2pt}V(x)\in\mathcal{R}_n,\hspace{2pt}\gamma\in\mathbb{R}^+}\beta\nonumber\\[2pt]	
	&\text{Subject to}:\nonumber\\[2pt]
	&-(V(x)-\gamma)+(p(x)-\beta)s_{10}\in\mathcal\sum_n\label{19}
\end{align}
Here, $V(x)$ is present in equation \eqref{19} and we know that $V(x)$ needs to satisfy \eqref{11a} and \eqref{11b}. So, combining \eqref{11a}, \eqref{11b} and \eqref{19}, we can write,
\begin{subequations}
	\begin{align}
		&\beta^*=\max_{\mathbf{s_0},\hspace{2pt}\mathbf{s_{10}\in\mathcal\sum_n},\hspace{2pt}V(x)\hspace{2pt}\in\hspace{2pt}\mathcal{R}_n,\hspace{2pt}\gamma\hspace{2pt}\in\hspace{2pt}\mathbb{R}^+}\beta\nonumber\\[2pt]	
		&\text{Subject to}:
		V(x)-l_1\in\mathcal{\sum}_n\label{20a}\\[2pt]
		-&\bigg[\frac{\partial V}{\partial x}f(x)+l_2\bigg]+(V(x)-\gamma)s_0\in\mathcal{\sum}_n\label{20b}\\[2pt]
		-&(V(x)-\gamma)+(p(x)-\beta)s_{10}\in\mathcal\sum_n\label{20c}
	\end{align}\label{20}
	Eq. \eqref{20} depicts the process of maximizing the variable $\beta$. However, in doing so, the optimal values of $V$ and $\gamma$ are also obtained. It is noted that incorporating an SF does not alter the existing equation represented by \eqref{11}, but rather it imposes an additional constraint. The utilization of a single SF in conjunction with Eq. \eqref{11} yields superior results compared to using Eq. \eqref{11} alone. However, Eq. \eqref{20} only estimates a portion of the actual ROA, particularly for systems with non-symmetric or unbounded actual ROA. This is discussed and illustrated through simulations of the Van der Pol system in the simulation section.  In the subsequent section, we propose a novel method to overcome this limitation.
\end{subequations}
\section{Union Theorem With Its Application to Estimate the Polynomial LF and Related ROA}
\begin{figure}[H]
	\begin{center}
		\includegraphics[scale=.5]{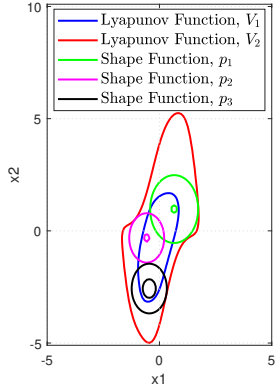}
		\caption{Union Theorem}\label{4ms}
	\end{center}
\end{figure}

Eq. \eqref{20c} expresses that the enclosed region defined by the SF should always be inside the region defined by the LF and $\gamma$. The SOSOPT program increases the value of $\beta$ in each iteration to enclose more regions under the SF and then generates a new LF for a fixed $\gamma$, which encircles the SF's region. This iteration method is known as V-S iteration and will be discussed in a later section.\\[3pt]
To estimate the ROA based on a single SF in SOS optimization (as expressed by \eqref{20}), the V-S iteration algorithm uses an increased number of iterations to enlarge the ROA estimation. However, this is not a universal remedy. For example, in cases of early converged optimization and/or numerical in-feasibility, particularly for a system with an unbounded or irregular ROA, this approach may not be effective. In addition to that, the geometric center of the SF, either fixed or adaptive, being located at the origin, limits the estimation, especially for non-symmetric or unbounded ROA. The reason is that after some iterations, the SF will be very close to the boundary of a non-symmetric ROA. Further increments of $\beta$ will not give any new surrounded LF as it is about to touch the boundary.
The estimated ROA might be very close to the real ROA at some regions of the boundary, however, at the same time, it stops enlargement in other regions.
This problem will be further demonstrated and discussed in the simulation section using the Van der Pol system.\\[3pt]
The idea that irregular ROA could be better determined by a combination of different SFs with shifted centers was proposed in \cite{r23}. However, in that paper, the authors used R-composition,\cite{r24} to obtain the final LF and ROA through Boolean operations that ultimately provided a non-polynomial LF, which can not be used to design the controller and construct the CLF using the SOS optimization technique as it works only with polynomial function. For these reasons, we proposed Union Theorem and developed a method using it in the SOS optimization to estimate the ROA. This method addresses both the issues related to the single SF-based approach and the R-composition-based method, as the estimated LF is polynomial and can capture areas in other directions if one direction becomes saturated.\\[3pt]
The basic idea is the introduction of multiple SFs in different places and then forcing the LF to encircle all of them. In every iteration, the value of all $\beta$ of all SFs will increase, and the new LF will capture more regions by enclosing them. The concept has been graphically shown in Fig.~\ref{4ms}, where LFs, $V_1$ and $V_2$ encircled the SFs, $p_1$, $p_2$ and $p_3$ and their increment, respectively.
We have the following Theorem and its proof based on the proposed concept.\vspace{4pt}
\begin{theorem}[Union Theorem]
	Let us consider that the given polynomials $V$, $p_1$,$...$,$p_n$ define sets $A_V$, $A_1$,$...$,$A_n$ such that,
	\begin{align*}
		A_V &= \{x\in \mathbb{R}^n: V\leq0\}\\[2pt]	
		A_1 &= \{x^1\in \mathbb{R}^n: p_1\leq0\}\\[-10pt]
		&\hspace{5pt}\vdots\\[-7pt]
		A_n &= \{x^n\in \mathbb{R}^n: p_n\leq0\}
	\end{align*}
	Now, if there exist polynomials $s_1\geq 0$,$...$,$s_n\geq 0$ such that,
	\begin{subequations}
		\begin{align}
			&-V+p_1s_1\geq 0\\[-10pt]
			&\hspace{25pt}\vdots\nonumber\\[-6pt]
			&-V+p_ns_n\geq 0\tag{21n}	
		\end{align}
	\end{subequations}
	Then, $A_1\cup A_2\cup ...\cup A_n\subseteq A_V$.
\end{theorem}\vspace{4pt}
\emph{\textbf{Proof:}} 
Let us consider that there exist polynomials $s_1\geq 0$,$...$,$s_n\geq 0$, such that,
\begin{subequations}
	\begin{align}
		&-V+p_1s_1\geq 0\label{22a}\\[-10pt]
		&\hspace{25pt}\vdots\nonumber\\[-6pt]
		&-V+p_ns_n\geq 0\tag{22n}	
	\end{align}
\end{subequations}
\begin{align*}
	\text{From \eqref{22a}, we get,}\hspace{5pt}	
	&-V+p_1s_1\geq 0\\[2pt]
	\Rightarrow &V\leq p_1s_1
\end{align*}
From the definition of the set $A_1$, we get,
\begin{align*}
	&x^1\in A_1\Rightarrow p_1\leq 0\\[2pt]
	\Rightarrow &p_1s_1\leq 0
	\hspace{5pt}\Big[\text{as}\hspace{3pt} s_1\geq 0\Big]\hspace{5pt}\forall\hspace{5pt}x^1\\[2pt]	
	&V\leq p_1s_1\leq 0\hspace{5pt}\forall\hspace{5pt}x^1\\[2pt]
	&V\leq 0\hspace{5pt}\forall\hspace{5pt}x^1
\end{align*}
Here $x^1\in A_1$ satisfies both $V\leq 0$ and $p_1\leq 0$. So, $x^1\in A_V$ also, which means $A_1\subseteq A_V$. Similar way, we can prove that $A_2\subseteq A_V$,$...$,$A_n\subseteq A_V$.
\begin{center}
	$\therefore$\hspace{5pt}$A_1\cup A_2\cup ...\cup A_n\subseteq A_V$
\end{center}
The Theorem also can be presented in the following way,\vspace{2pt}
\begin{theorem}[Union Theorem]	
	Let us consider that the given polynomials $V$, $p_1$,$...$,$p_n$ define sets $A_V$, $A_1$,$...$,$A_n$ such that,\vspace{0.1pt}
	\begin{align*}
		A_V &= \{x\in \mathbb{R}^n: V-\gamma\leq0\}\\[2pt]	
		A_1 &= \{x^1\in \mathbb{R}^n: p_1-\beta_1\leq0\}\\[-10pt]
		&\hspace{5pt}\vdots\\[-7pt]
		A_n &= \{x^n\in \mathbb{R}^n: p_n-\beta_n\leq0\}
	\end{align*}
	Here, $\Big\{\gamma, \beta_1,...\beta_n\Big\}\in\mathbb{R}^+$. Now if there exit polynomials
	$s_1\in\mathcal{\sum}_n$,$...$,$s_n\in\mathcal{\sum}_n$, such that,
	\begin{subequations}
		\begin{align}
			&-(V-\gamma)+(p_1-\beta_1)s_1\in\mathcal{\sum}_n\\[-10pt]
			&\hspace{25pt}\vdots\nonumber\\[-7pt]
			&-(V-\gamma)+(p_n-\beta_n)s_n\in\mathcal{\sum}_n\tag{23n}	
		\end{align}\label{23}
	\end{subequations}
	Then, $A_1\cup A_2\cup ...\cup A_n\subseteq A_V$.
\end{theorem}\vspace{4pt}
\emph{Union Theorem} tells us that we can obtain a $V$ that will encircle the union of areas generated by all the SFs, ${p_i}$. It is one kind of S-procedure like \emph{Theorem 3}. \emph{Theorem 3} provides a function to encircle the intersection regions, while \emph{Union Theorem} provides a function to encircle the union of regions defined by other positive functions.\\[3pt]
If we take $n$ SFs in the system and apply the \emph{Union Theorem}, then Eq. \eqref{20} can be written in the following way,\vspace{0.1pt}
\begin{subequations}
	\begin{align}
		&\Big[\beta^*_1,...,\beta^*_n\Big]=\max_{\mathbf{s_{0}},\hspace{2pt}\mathbf{s_{1},...,s_{n}\in\mathcal\sum_n},\hspace{2pt}V(x)\in\mathcal{R}_n,\hspace{2pt}\gamma\in\mathbb{R}^+}\beta_1,...,\beta_n\nonumber\\[2pt]	
		&\text{Subject to}:\nonumber\\[2pt]
		&\hspace{12pt}V-l_1\in\mathcal{\sum}_n\label{24a}\\[2pt]
		&-\bigg[\frac{\partial V}{\partial x}f+l_2\bigg]+(V-\gamma)s_0\in\mathcal{\sum}_n\label{24b}\\[2pt]
		&-(V-\gamma)+(p_1-\beta_1)s_1\in\mathcal{\sum}_n\label{24c}\\[-10pt]
		&\hspace{25pt}\vdots\nonumber\\[-7pt]
		&-(V-\gamma)+(p_n-\beta_n)s_n\in\mathcal{\sum}_n\tag{24n}\label{24n}
	\end{align}\label{24}
\end{subequations}
\textbf{Remark 1.} Eq. \eqref{24a} ensures the positive definiteness of $V$, which is the 1st condition for a LF.\\[3pt] 
\textbf{Remark 2.} Eq. \eqref{24b} tells us that the region enclosed by $V\leq\gamma$ should be always inside the stable region as $\dot{V}\leq-l_2$.\\[3pt] 
\textbf{Remark 3.} Eq. \eqref{24c}-\eqref{24n} make sure that the regions enclosed by $p_1\leq\beta_1,...,p_n\leq\beta_1$ should be always inside the region enclosed by $V\leq\gamma$.\vspace{-5pt}
\subsection*{\textbf{Benefits of the Proposed Method:}}
\begin{itemize}  
	\item {The proposed method estimates a very large subset of the ROA, with a polynomial LF, unlike the RcomSSF, where the final LF is not a polynomial, which cannot be used in the SOSOPT Toolbox (the whole P-staz idea is based on the polynomial function) as the Toolbox only works with polynomial functions.}\vspace{3pt} 
	\item {The method will be very efficient for designing the controller and constructing the CLF with enhanced ROA, as the obtained polynomial LF can be fed into the system's algorithm.}
\end{itemize}
\section{Numerical Algorithm}
 To implement the idea, proposed by the \emph{Union theorem}, the efficient numerical algorithm is elaborated in this paper.\\[3pt]
 Eq. \eqref{24} is a bi-linear optimization problem as $V$ and $s_i$ are coupled and both are linear in the equations. The problem is solved using a two-way iterative search algorithm, known as $V-s$ iteration. The details of this algorithm can be found in \cite{r17} and \cite{r18}. We have used the SOSOPT Toolbox \cite{r25} to solve Eq. \eqref{24}. The toolbox with MATLAB code is available in \cite{r9}. The $V-s$ algorithm brakes Eq. \eqref{24} into three steps to get the ultimate result. The first two steps named $\gamma$-step, $\beta$-step involve optimizing $\gamma$ and $\beta_i$, respectively, while the third step named $V$-step is a feasibility problem that finds a new LF $V$ that satisfies all constraints under it. We modified the original $V-s$ algorithm to include multiple SFs in the optimization procedure according to Eqs. \eqref{24c}-\eqref{24n}. The resultant algorithm is as follows.
\subsection*{\textbf{V-s Iteration Algorithm:}}
\textbf{1.} \textbf{$\gamma$-step:} Solve for $s_0$ and $\gamma^*$ for a given $V$ and fixed $l_2:$
\begin{align}
	&\gamma^*=\max_{\mathbf{s_{0}}\in\mathcal\sum_n}\gamma\hspace{6pt}	
	\text{s.t.}:-\bigg[\frac{\partial V}{\partial x}f+l_2\bigg]+(V-\gamma)s_0\in\mathcal{\sum}_n\label{25}
\end{align}
\begin{subequations}
	\textbf{2.} \textbf{$\beta$-step:} Solve for $s_{1},...,s_{n}$ and $\beta_1^*,...,\beta_n^*$ for a given $V$, $p_i$ and using obtained $\gamma^*$ from $\gamma$-step:
	\begin{align}
		\Big[\beta^*_1,...,&\beta^*_n\Big]=\max_{\hspace{2pt}\mathbf{s_{1},...,s_{n}\in\mathcal\sum_n}}\beta_1,...,\beta_n\hspace{6pt}	
		\text{s.t.}:\nonumber\\
		&-(V-\gamma^*)+(p_1-\beta_1)s_1\in\mathcal{\sum}_n\label{26a}\\[-10pt]
		&\hspace{25pt}\vdots\nonumber\\[-7pt]
		&-(V-\gamma^*)+(p_n-\beta_n)s_n\in\mathcal{\sum}_n\tag{26n}\label{26n}
	\end{align}\label{26}
\end{subequations}
\begin{subequations}
	\textbf{3.} \textbf{$V$-step:} Using the obtained $\gamma^*$, $s_{0},s_{1},...,s_{n}$ and $\beta_1^*,...,\beta_n^*$ from previous two steps, solve for a new $V$ which satisfies the following : 
	\begin{align}
		&\hspace{12pt}V-l_1\in\mathcal{\sum}_n\label{27a}\\[2pt]
		&-\bigg[\frac{\partial V}{\partial x}f+l_2\bigg]+(V-\gamma^*)s_0\in\mathcal{\sum}_n\label{27b}\\[2pt]
		&-(V-\gamma^*)+(p_1-\beta_1^*)s_1\in\mathcal{\sum}_n\label{27c}\\[-10pt]
		&\hspace{25pt}\vdots\nonumber\\[-7pt]
		&-(V-\gamma^*)+(p_n-\beta_n^*)s_n\in\mathcal{\sum}_n\tag{27n}\label{27n}
	\end{align}\label{27}
\end{subequations}
\textbf{4.} \textbf{Scale $V$:} Replace $V$ with $V/\gamma^*$ after each $V$-step. This scaling process roughly normalizes $V$ and tends to keep the $\gamma^*$ computed in the next step ($\gamma$-step) close to unity.\\[3pt]
\textbf{5. Repeat} all the steps from $1-4$ using the scaled $V$ from step 4 to the $\gamma$-step as an input. Continue the whole process until the final feasible LF $V/\gamma^*$ is obtained, which gives the final ROA in $\Omega^*:=\{x\in\mathbb{R}^n:V^*(x)\leq1\}$.\vspace{3pt}\\
\emph{\textbf{Stopping Criteria:}} \\[3pt] 
\textbf{A.} Numerical Infeasibility:- The algorithm stops if it is not possible to obtain any new LF that satisfies all the constraints of $V$-step.\\[3pt]
\textbf{B.} Negligently Increment of $\beta$:- The algorithm stops if any two consecutive estimations of all $\beta_i$ are less than a pre-defined tolerance value. This criteria can be used if the SF is fixed.\\[3pt]
\textbf{C.} Fixed Iteration Number:- The algorithm comes to an end after running through all the specified iterations.\\[3pt]
The algorithm could stop for other reasons, such as if $\beta_i$ and $s_i$ values are not found. This can be fixed by relocating the SF's center or changing the SF.\vspace{3pt}\\
\textbf{Remark 1.} Let us consider the $\gamma$-step. The central idea is that it takes a fixed value for $\gamma$ and then finds an SOS polynomial $s_0$ so that the constraint Eq. \eqref{25} becomes an SOS polynomial. $\gamma$ is then increased, and a new $s_0$ is located to achieve the same transformation. This is repeated until no additional $s_0$ can be found for a fixed $\gamma$. In order to obtain an SOS polynomial $s_0$ that makes Eq. \eqref{25} an SOS polynomial for a given $\gamma$, the below procedure described in \cite{r25} is followed.
\begin{align}
	-\bigg[\frac{\partial V}{\partial x}f+l_2\bigg]+(V-\gamma)s_0=C(x)
\end{align}
Here $C(x)$ is an SOS polynomial. Let us consider,
$s_0(x)=Z_0'(x)Q_0Z_0(x)$ and $C(x)=Z_C'(x)Q_CZ_C(x)$, where $Z_0$ and $Z_C$ are vector of monomials, while $Q_0$ and $Q_C$ are symmetric Gram matrices and $Q_0\succeq0$ and $Q_C\succeq0$ as $s_0$ and $C$ are SOS polynomials. So, we can write,
\begin{align}
	-\bigg[\frac{\partial V}{\partial x}f+l_2\bigg]+(V-\gamma)Z_0'(x)Q_0Z_0(x)\nonumber\\
	=Z_C'(x)Q_CZ_C(x)
\end{align}
Equating the coefficient leads to linear equality constraints where the elements of $Q_0$ and $Q_C$ appear as variables. There exist matrices $A_d$, $A_q$ and a vector $b$ such that these equality constraints can be represented as $A_dd+A_qq=b$ where $d:=vec(Q_0)$ and $q:=vec(Q_q)$. So, Eq. \eqref{25} can be written,
\begin{align}
	&\text{Given matrices $A_d$ and $A_q$ and a vector $b$,}\nonumber\\
	&\text{Find $Q_0\succeq0$ and $Q_C\succeq0$  s.t. $A_dd+A_qq=b$}
\end{align}
This is a semi-definite programming problem and can be solved using SeDuMi \cite{r26, r27} . The $\beta$-step and $V$-step can also be solved similarly.\\[3pt]
\textbf{Remark 2.} The following degrees are required to satisfy for a feasible solution of \eqref{25}-\eqref{27}, (\cite{r19}, \cite{r20}, \cite{r23}, \cite{r30}),
\begin{align*}
	&\text{deg}(V)\geq \text{deg}(l_1)\\[4pt]	
	&\text{deg}(p_i)+\text{deg}(s_i)\geq \text{deg}(V),\hspace{5pt}[i=1,...,n]\\	
	&\text{deg}(V)+\text{deg}(s_0)\geq \text{max\{deg($\frac{\partial V}{\partial x}f$),\hspace{2pt} deg($l_2$)\}}
\end{align*}
\textbf{Remark 3.}  An initial LF $V_0$ is required in the $\gamma$-step to use the $V-s$ algorithm. It can be obtained in various ways, for example, from well-known Lyapunov equation (LE).
\begin{subequations}
	\noindent\begin{minipage}{.5\linewidth}
		\begin{align}
			V_0=x^TPx\label{31a}
		\end{align}
	\end{minipage}%
	\begin{minipage}{.5\linewidth}
		\begin{align}
			A^TP+PA=-Q\label{31b}
		\end{align}
	\end{minipage}\label{31}
\end{subequations}
Where,
\begin{align*}
	A=\Big(\frac{\partial f}{\partial x}\Big)_{x=0},\hspace{4pt} \text{P is a positive definite matrix and $Q>0$.}
\end{align*}
The solution of the LE, \eqref{31b} gives $P$ and substituting $P$ in \eqref{31a} provides the initial LF $V_0$. During the first iteration, $V_0$ is fed into the $\gamma$-step of the $V-s$ iteration, sequentially executing the $\gamma$, $\beta$, and $V$ steps, resulting in a new LF in the $V$ step. In the next iteration, the algorithm goes back to the $\gamma$-step again and uses the normalized new LF as an initial LF. and the process continues until the stopping criteria are met.\\[3pt]
\textbf{Remark 4.}
Equation \eqref{26} states that the SF should always be within the LF used in the $\gamma$-step. According to \cite{r23}, multiple rounds of iterations can improve the estimation of the ROA. For example, after running the algorithm, the final LF $V_{1R}$ and a larger value of $\beta$ of the SF $p_{1R}$ are obtained. However, $V_{1R}$ may not fully cover the ROA, even with multiple SFs. To capture a larger area, the algorithm is run again with $V_{1R}$ as the initial LF for the second round of iteration. Different types of SFs may be used in each round, and sometimes using different types can lead to better results. However, it is noted that the LF may become saturated after a certain round of iterations and cannot capture more area, even with additional rounds. It may also be the case that no new $V$ that satisfies constraint \eqref{27} can be found, despite there being a significant stable region outside the captured region of $V$.
\section{Selection of Shape Function}
The SF is crucial for getting a better estimation of the ROA. The total iteration time also depends on the SF. Different types of SFs will give different LFs with varying ROAs and total elapsed times. According to the definition, any positive definite function can be used as an SF. An origin-centered quadratic function is frequently used as it is computationally cost-efficient \vspace{-5pt}
\begin{align}
	p(x)=x^TNx.
\end{align}
The shape matrix $N$ is a positive definite matrix that influences the SF's shape, ROA size, and $V$ calculation. $N$ selection is a case-specific, with $N=I$ (Identity Matrix) being a standard choice without prior knowledge.\vspace{3pt}\\
Alternatively, an adaptive method for selecting the origin-centered SF is proposed in \cite{r30}, where a new SF is generated each iteration by utilizing the quadratic portion of the newly estimated LF as the SF, which then replaces the previous SF.\vspace{3pt}\\
Remark 4 indicates the need for multiple rounds of iterations to obtain an improved ROA estimate. Thus, an origin-centered SF alone is insufficient as it may approach the unstable region after some iterations and stop growing. A new type of SF, the shifted shape function, with a non-origin center, was introduced in \cite{r23} to address this problem. In our method, either an origin-centered or a shifted SF can be used in any iteration round, depending on the problem, with the constraint that the center is within the LF region obtained in the $\gamma$-step as per constraint \eqref{26}. The general shifted shape function with a shifting center $x^*$, proposed in \cite{r23}, is as follows
\begin{align}
	p(x)=(x-x^*)^TN(x-x^*).\label{33}
\end{align}
It is noted that a fixed $N$ may not yield good results, so we need to experiment with different $N$ in each iteration round and compare the estimated ROA to select the best one. To use the shifted shape function, as expressed in equation \eqref{33}, we need a shifting center $x^*$, which is calculated in the next section.
\subsection{Selection of shifting centre}
Choosing the correct shifting center is important for a better estimation of ROA. Various methods can be used for selecting the center. In this paper, we followed the guidelines  proposed in \cite{r23} and implemented them in the form of an efficient computational procedure. This procedure involves the following steps.\\[3pt]
\textbf{1.} Get a bounded LF ($V$ and related maximum $\gamma$)\\[3pt]
\textbf{2.} Consider a straight line from the origin with a pre-fixed inclination angle.\\[3pt]
\textbf{3.} Get an intersection point ($x^*_{I}$) on the boundary of the bounded LF by solving the equation of a straight line and the bounded LF.\\[3pt]
\textbf{4.} The shifting center can be chosen at $x^*=\rho_\alpha x^*_{I}$\\[3pt]
We have given the mathematical expression to choose the shifting center for the 2-D case only.
Let us consider the equations of bounded LF and 2-D straight line are,\vspace{-5pt}\\
\begin{subequations}
	\noindent\begin{minipage}{.5\linewidth}
		\begin{align}
			V(x_1,x_2)=\gamma^*\label{34a}
		\end{align}
	\end{minipage}%
	\begin{minipage}{.5\linewidth}
		\begin{align}
			x_2=\tan\theta x_1\label{34b}
		\end{align}
	\end{minipage}\label{34}\vspace{5pt}
\end{subequations}
Here, $\theta$ is the pre-defined inclination angle of the line, and for every shifting center, there is a fixed $\theta$. Consider the solution point or intersection point of the above equations to be ($x^*_{1I},x^*_{2I}$). The distance from the center to the point ($x^*_{1I}, x^*_{2I}$) is given by $\rho_\alpha=\sqrt{x^*_{1I}+x^*_{2I}}$. So, the shifting center is,\vspace{-5pt}\\
\begin{subequations}
	\noindent\begin{minipage}{.5\linewidth}
		\begin{align}
			x^*_1=\sigma\rho_\alpha\cos\theta\label{35a}
		\end{align}
	\end{minipage}%
	\begin{minipage}{.5\linewidth}
		\begin{align}
			x^*_2=\sigma\rho_\alpha\sin\theta\label{35b}	
		\end{align}
	\end{minipage}\label{35}\vspace{5pt}
\end{subequations}
Where, $\sigma\in(0,1)$. For the 3-D case, we have considered the following equation for the straight line,\vspace{-5pt}\\
\begin{subequations}
	\noindent\begin{minipage}{.35\linewidth}
		\begin{align}
			x_1=r\cos\theta\cos\psi\label{36a}
		\end{align}
	\end{minipage}%
	\begin{minipage}{.35\linewidth}
		\begin{align}
			x_2=r\cos\theta\sin\psi\label{36b}
		\end{align}
	\end{minipage}
	\begin{minipage}{.27\linewidth}
		\begin{align}
			x_3=r\sin\theta\label{36c}	
		\end{align}
	\end{minipage}\label{36}\vspace{5pt}
\end{subequations}
Here, $\theta$ and $\psi$ are the pre-defined elevation and azimuth angles. We can obtain the shifting center ($x^*_1,x^*_2,x^*_3$) for the 3-D case using the same procedure as in the 2-D case.\\[3pt]
Choosing $\sigma$ as either $0$ or $1$ is not viable since a value of $0$ does not provide a shifted shape function, and choosing $1$ places the center on the boundary, making it impossible to obtain any value for $\beta$ that satisfies the constraint \eqref{26}. The selection of $\sigma$ is crucial, with a smaller value resulting in the shifting center being closer to the origin, requiring more iterations to estimate the ROA beyond the parent level set. Conversely, a larger value of $\sigma$ would place the center closer to the boundary, giving a small initial expansion of $\beta$ and impeding proper estimation. Therefore, we must carefully choose $\sigma$ to achieve a larger ROA with fewer iterations. Centers closer to the convex boundary offer better estimation, as the concave direction estimation is already near the actual ROA, \cite{r23}. Furthermore, determining the optimal number of centers or SFs has no fixed rules and is problem-dependent, requiring an error and trial approach, as sometimes more centers provide better results at the cost of more computational power, while fewer centers may capture a larger region.\\[3pt]
In the next section, we present some examples and simulation results of the estimated ROA using the proposed method.  
\section{Simulation Results}
\begin{figure}[H]
	\centering
	\includegraphics[width=.7\columnwidth]{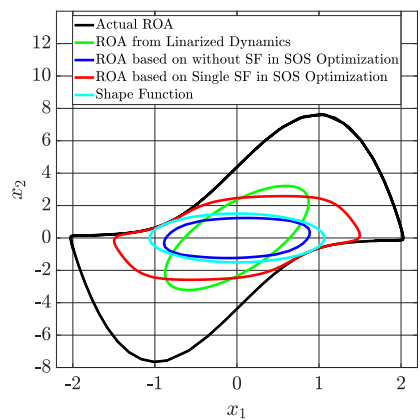}
	\caption{Estimated ROA}\label{6_f0}	
\end{figure}

\begin{figure*}[htbp]
	\centering
	\subfigure[Estimated ROA from 1st Round]{\includegraphics[scale=0.2]{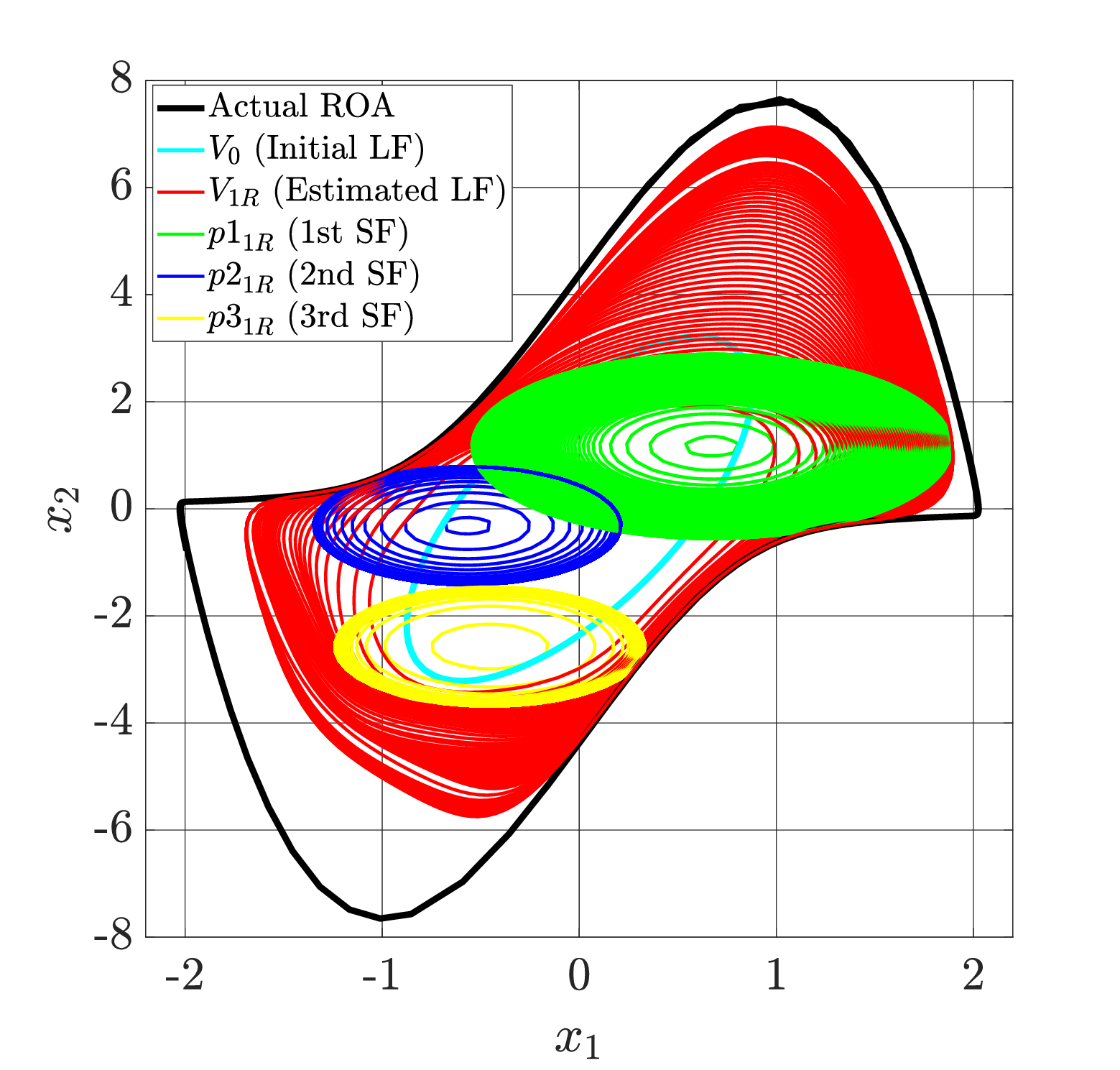}
		\label{5a_f}}
	\subfigure[Estimated ROA from 2nd Round]{\includegraphics[scale=0.2]{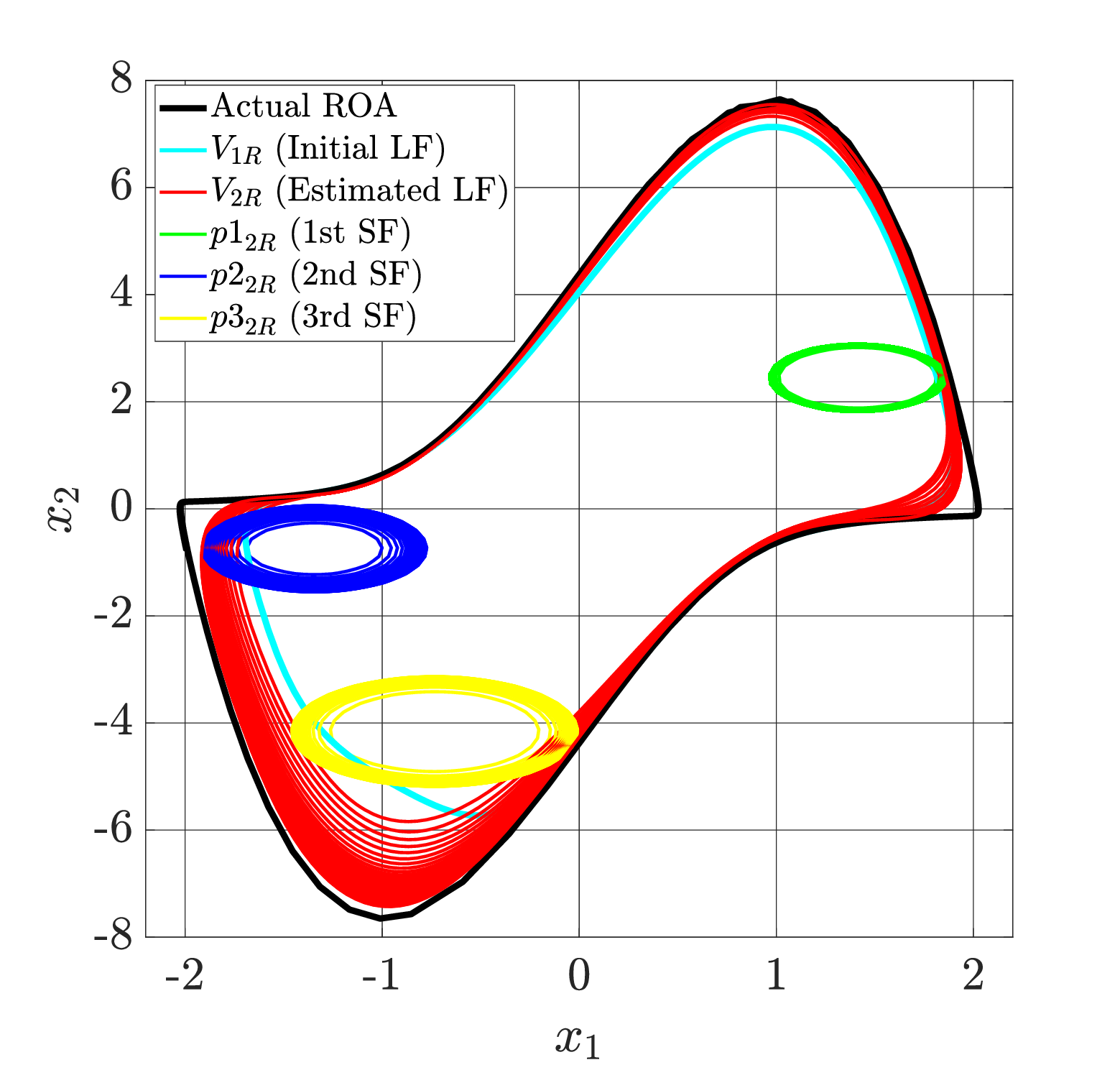}
		\label{5b_f}}	
	\subfigure[Estimated ROA from 3rd Round]{\includegraphics[scale=0.2]{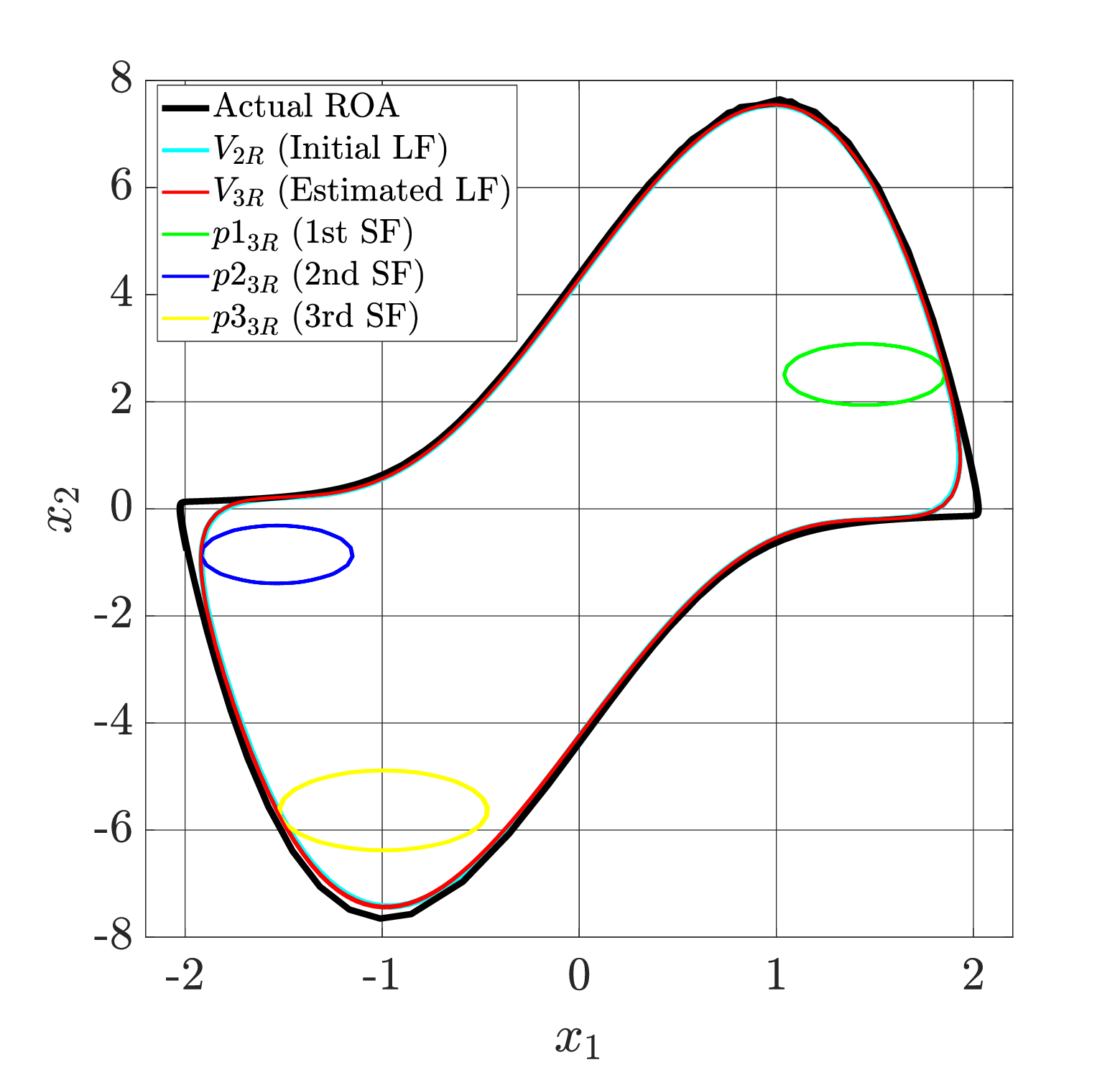}
		\label{5c_f}}\vfill
	\subfigure[Comparison Among Estimated ROAs from all Rounds]{\includegraphics[scale=0.2]{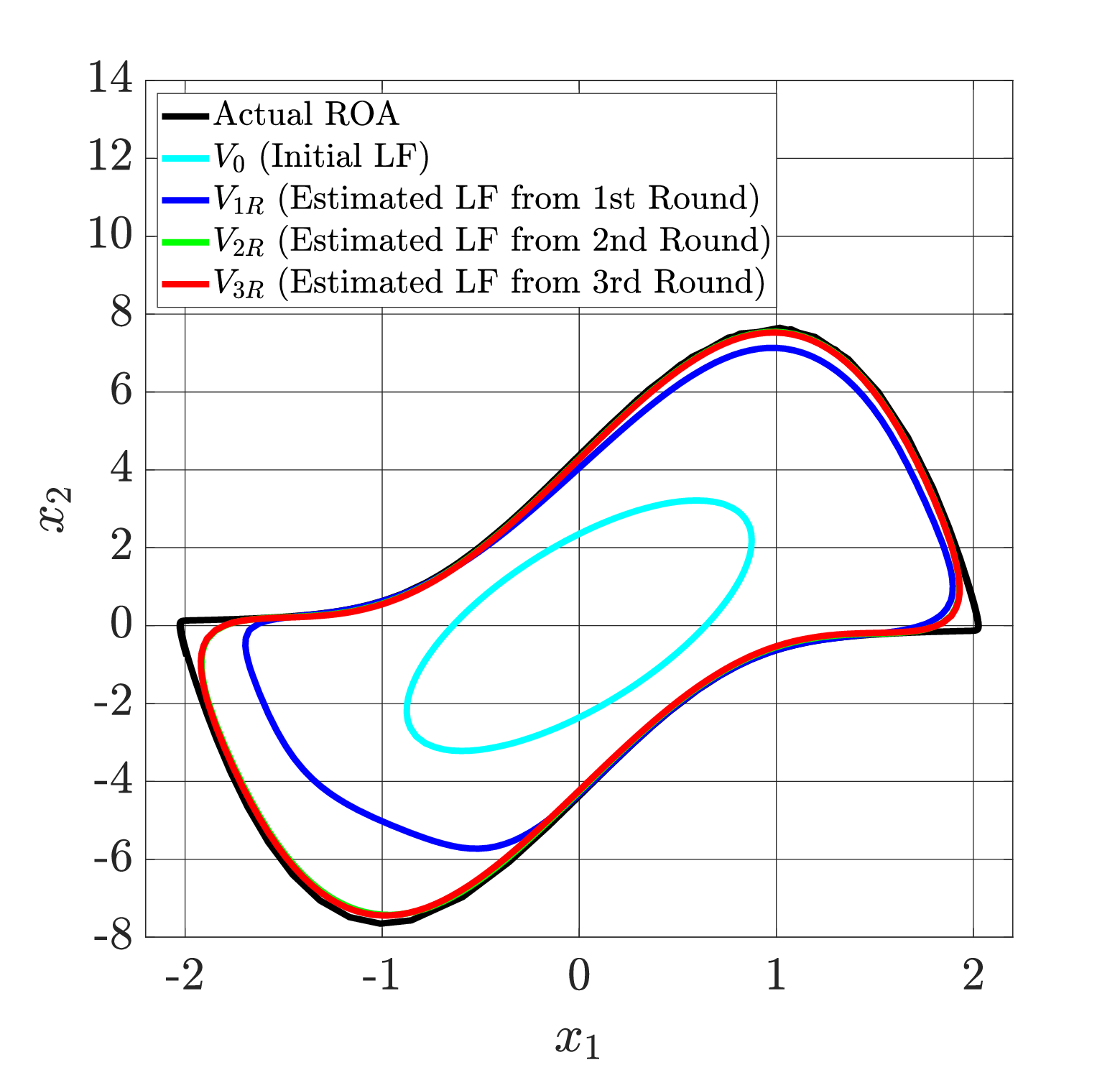}
		\label{5d_f}}
	\subfigure[Final Estimated ROA (Obtained from 3rd Round)]{\includegraphics[scale=0.2]{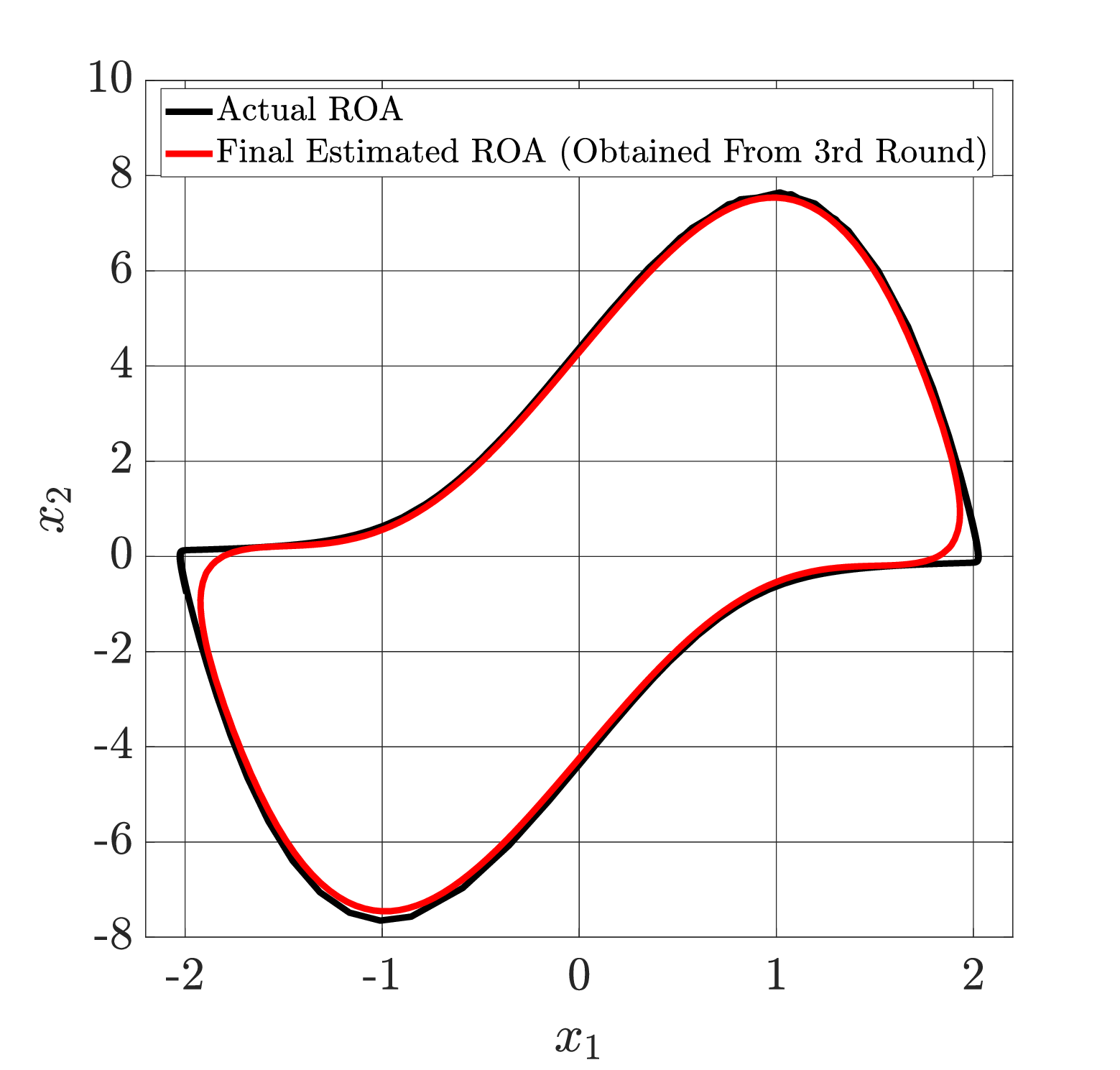}
		\label{5e_f}}
	\subfigure[Comparison Among Estimated ROAs based on Different Methods]{\includegraphics[scale=0.2]{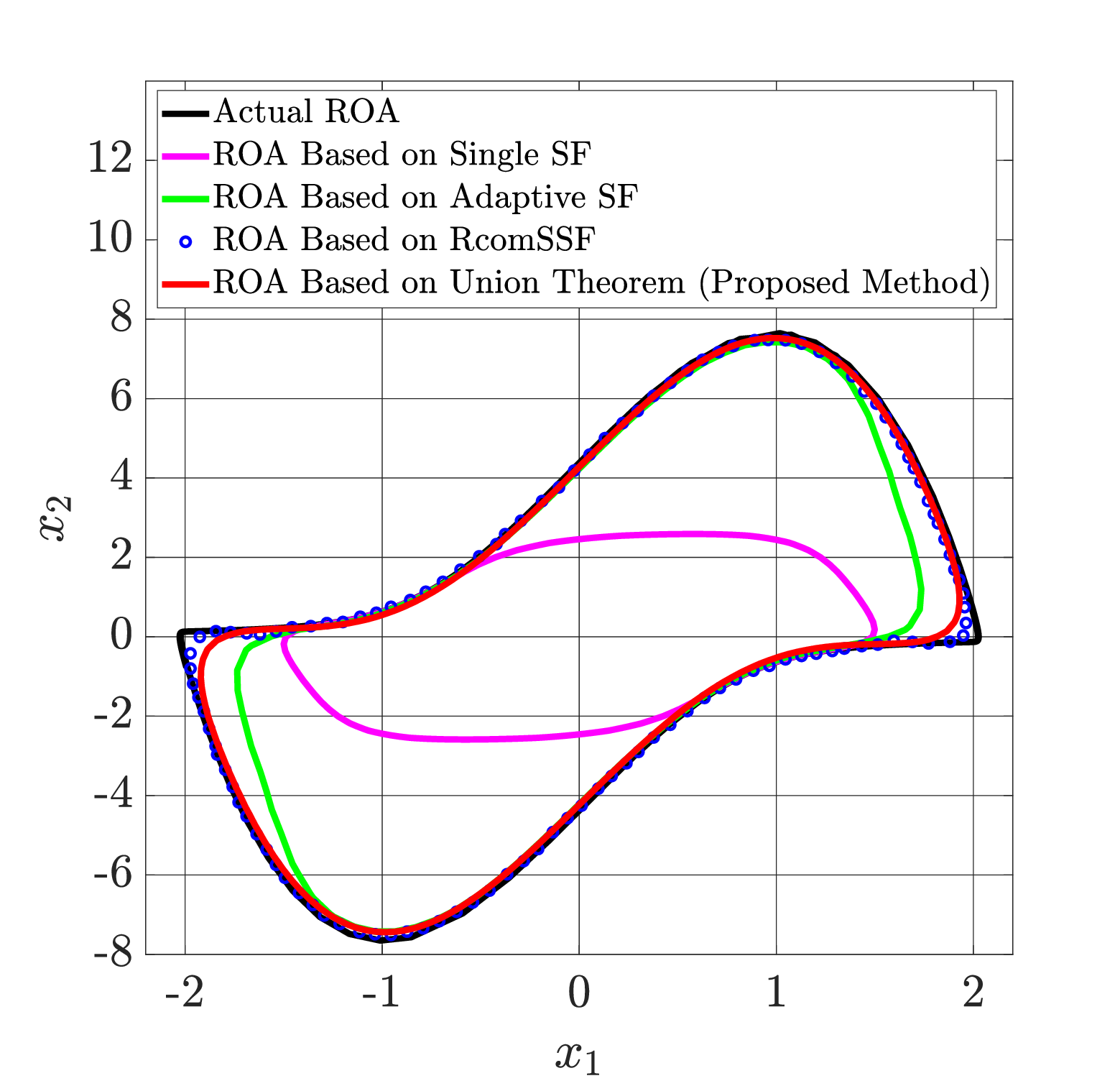}
		\label{5f_f}}
	\caption{ROA Estimation for Example 1}
	\label{5_f}\vfill	
	\subfigure[Estimated ROA from 1st Round]{\includegraphics[scale=0.2]{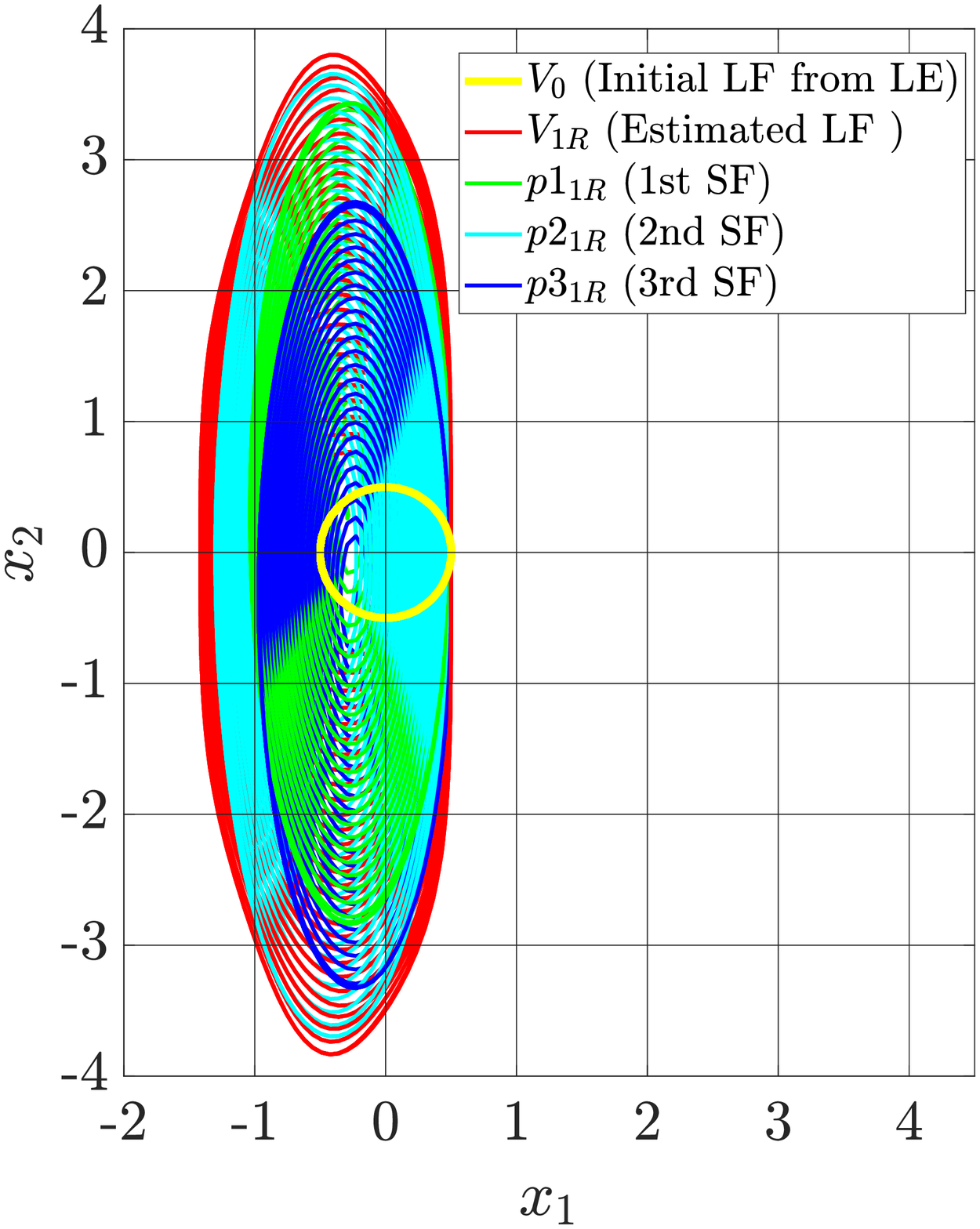}
		\label{6a_f}}
	\subfigure[Estimated ROA from 2nd Round]{\includegraphics[scale=0.2]{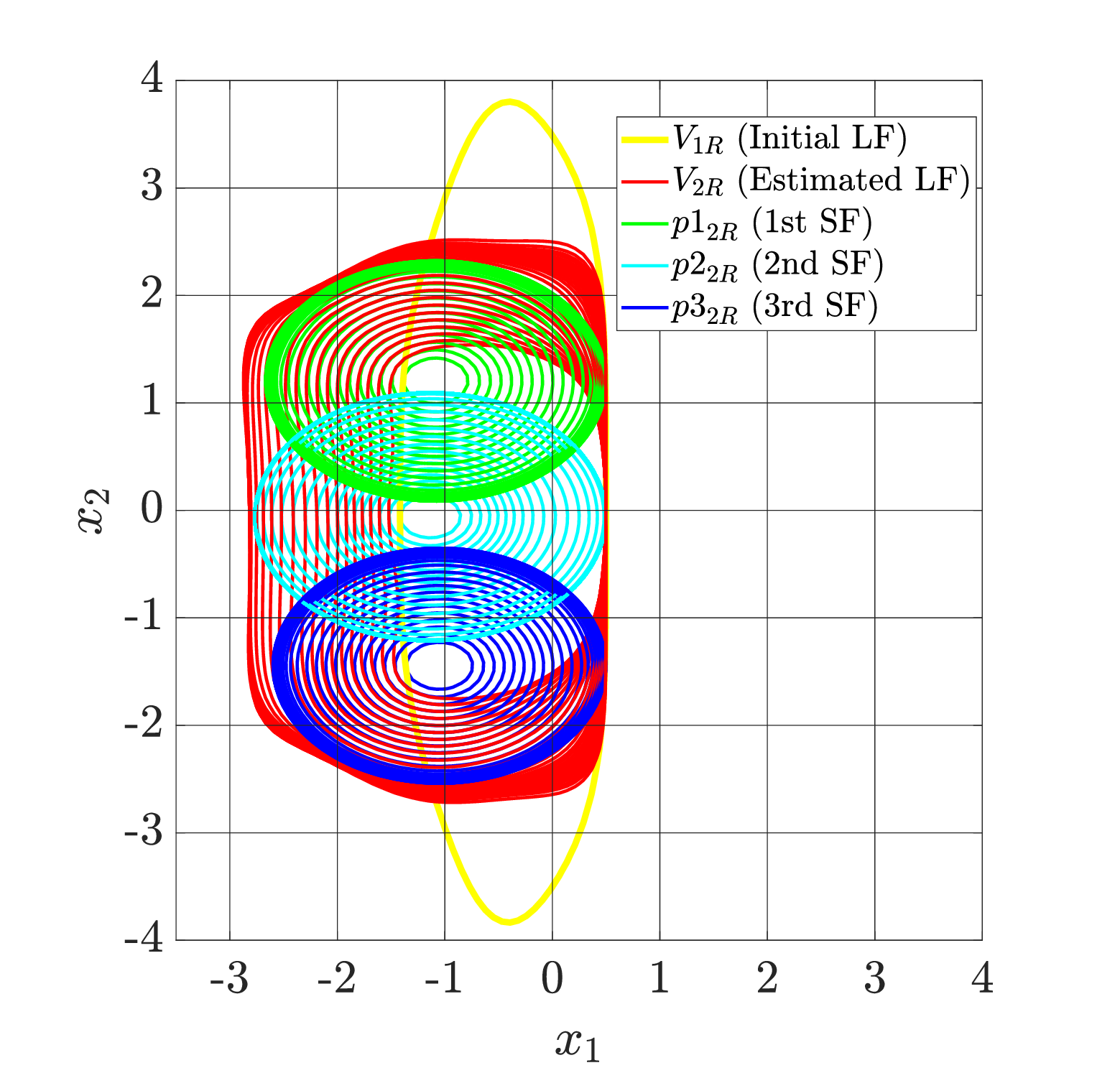}
		\label{6b_f}}
	\subfigure[Estimated ROA from 3rd Round]{\includegraphics[scale=0.2]{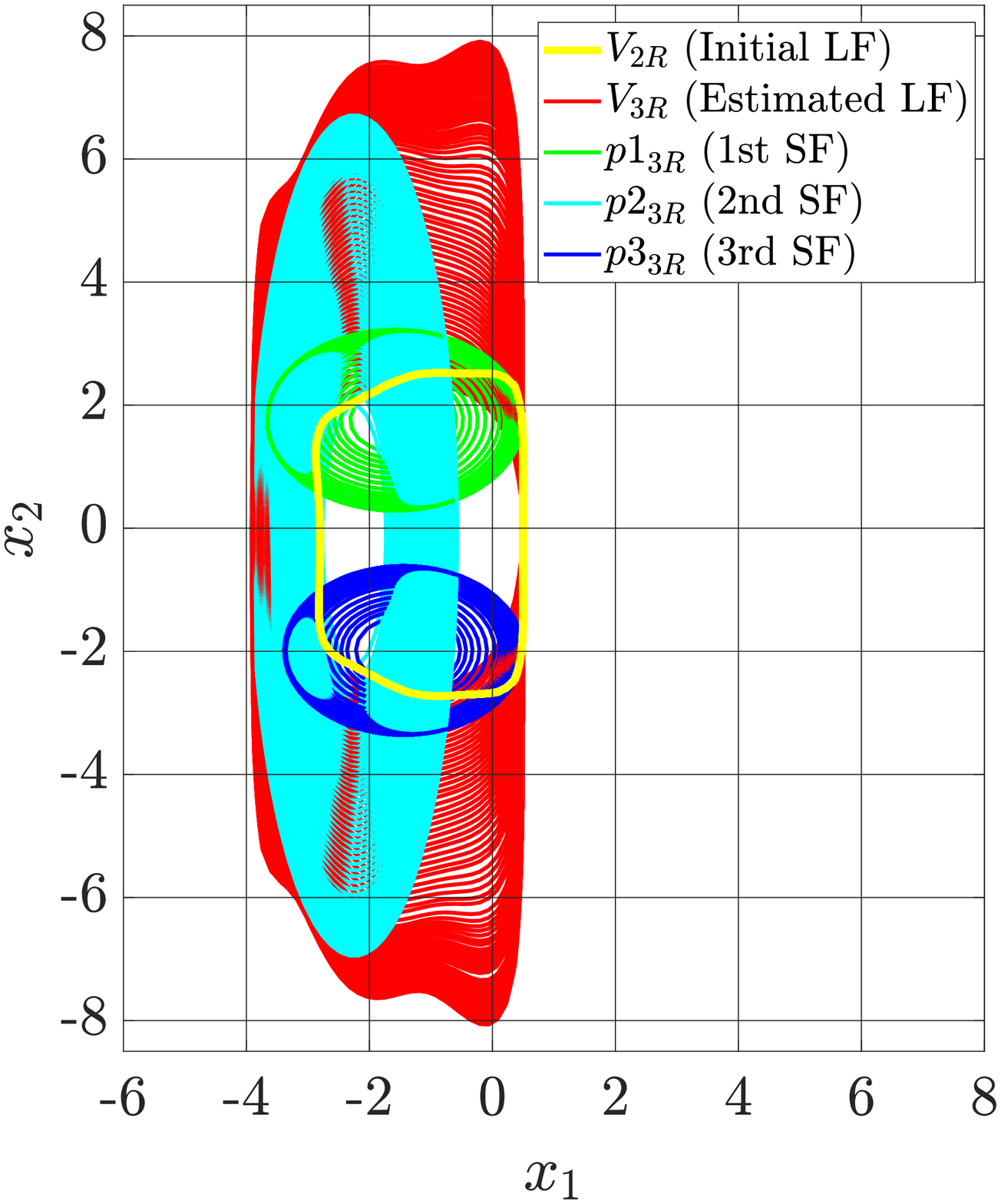}
		\label{6c_f}}\vfill
	\subfigure[Comparison Among Estimated ROAs from all Rounds]{\includegraphics[scale=0.2]{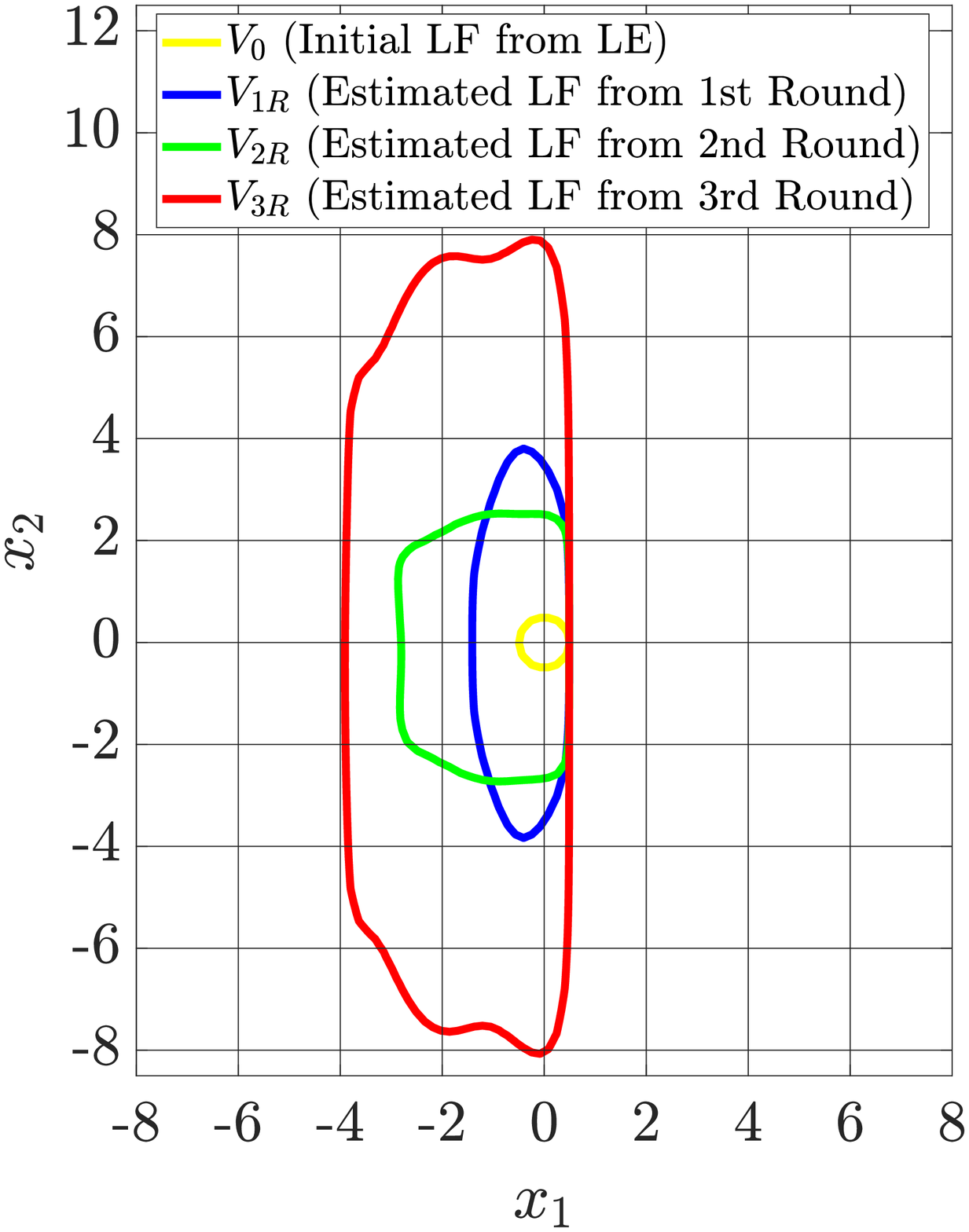}
		\label{6d_f}}
	\subfigure[Final Estimated ROA (Obtained from 3rd Round)]{\includegraphics[scale=0.2]{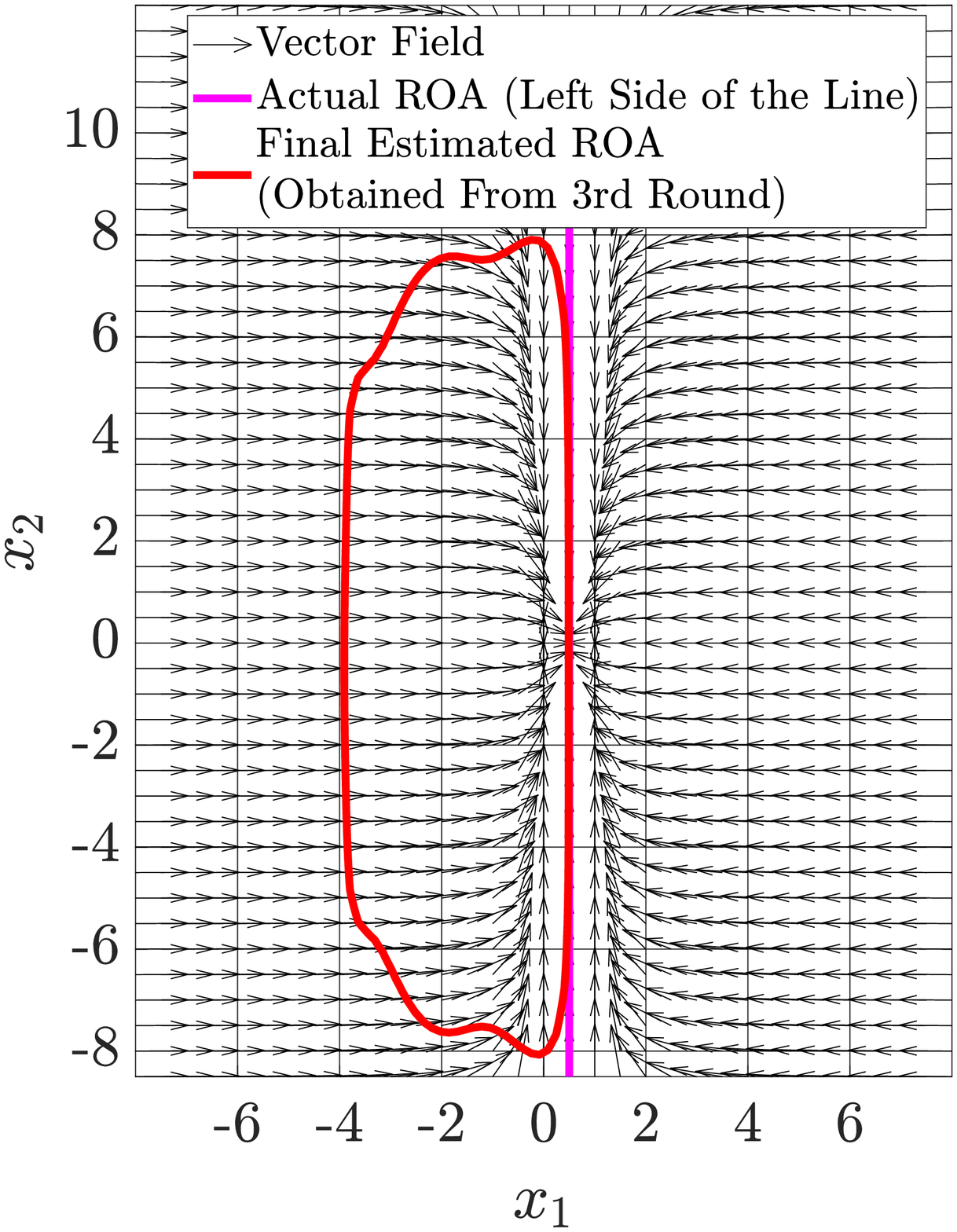}
		\label{6e_f}}
	\subfigure[Comparison Among Estimated ROAs based on Different Methods]{\includegraphics[scale=0.2]{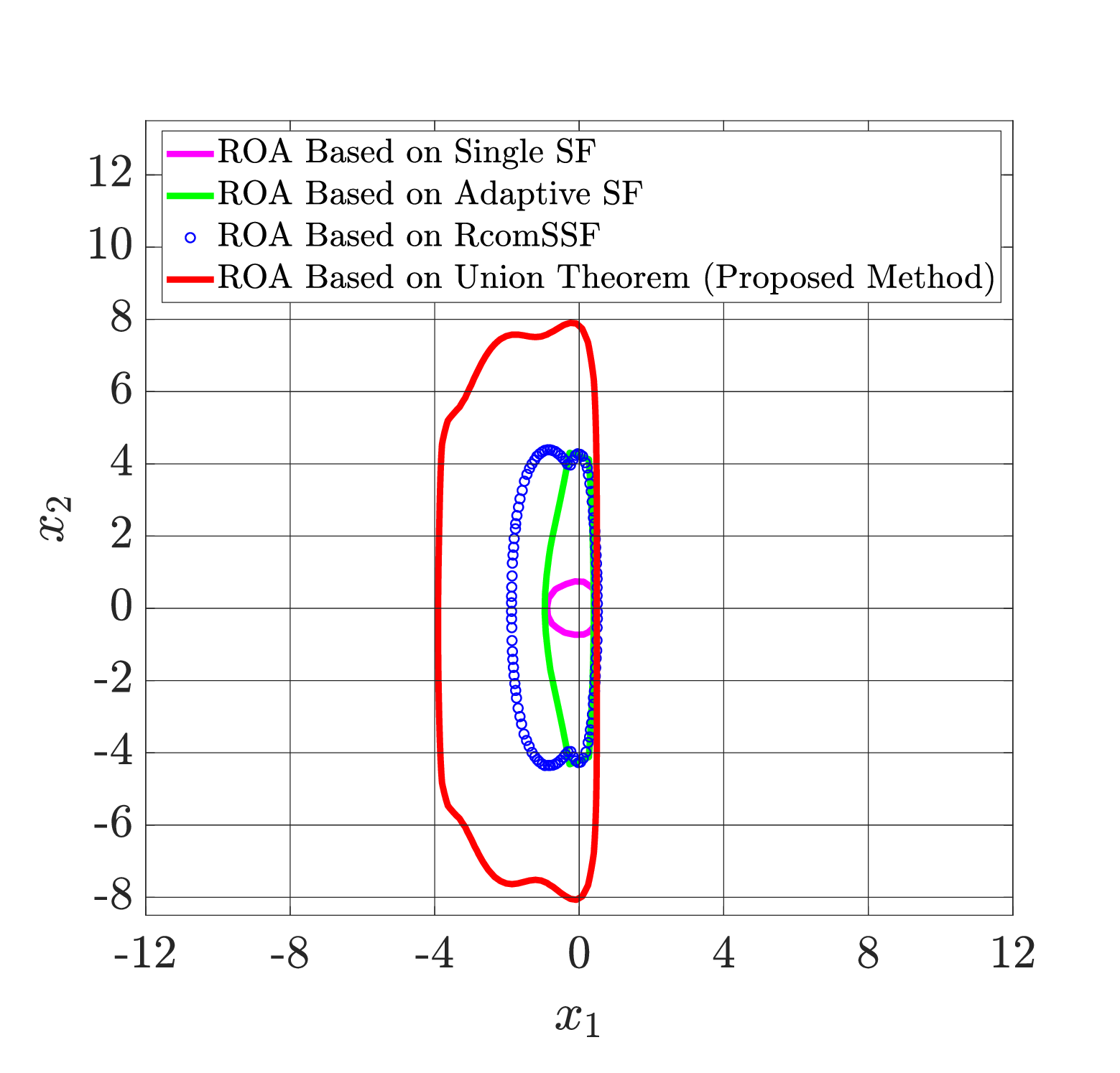}
		\label{6f_f}}
	\caption{ROA Estimation for Example 2}
	\label{6_f}
\end{figure*}

\begin{figure*}[htbp]
	\centering
	\subfigure[Estimated ROA from 1st Round]{\includegraphics[scale=0.2]{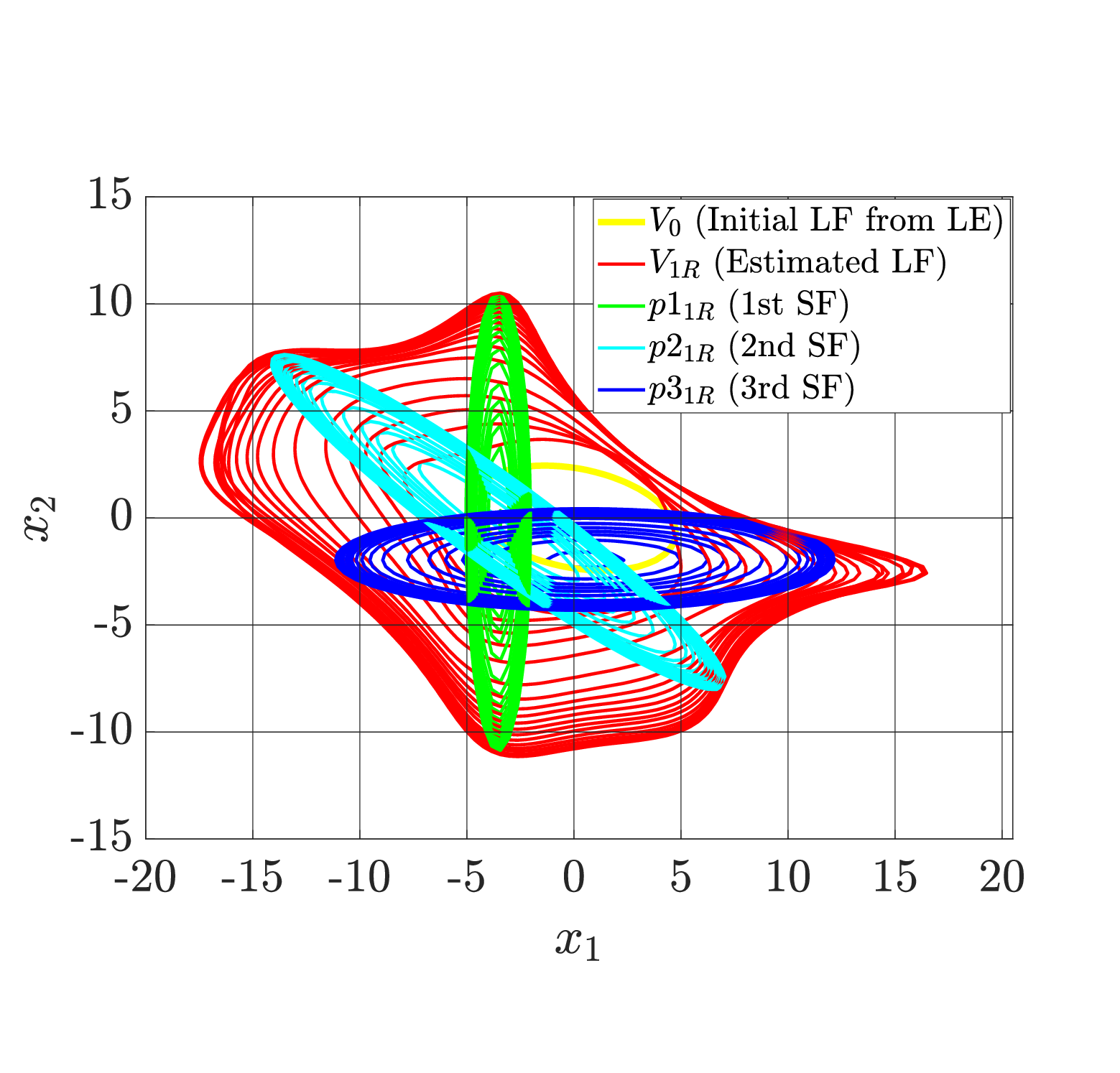}
		\label{7a_f}}
	\subfigure[Estimated ROA from 2nd Round]{\includegraphics[scale=0.2]{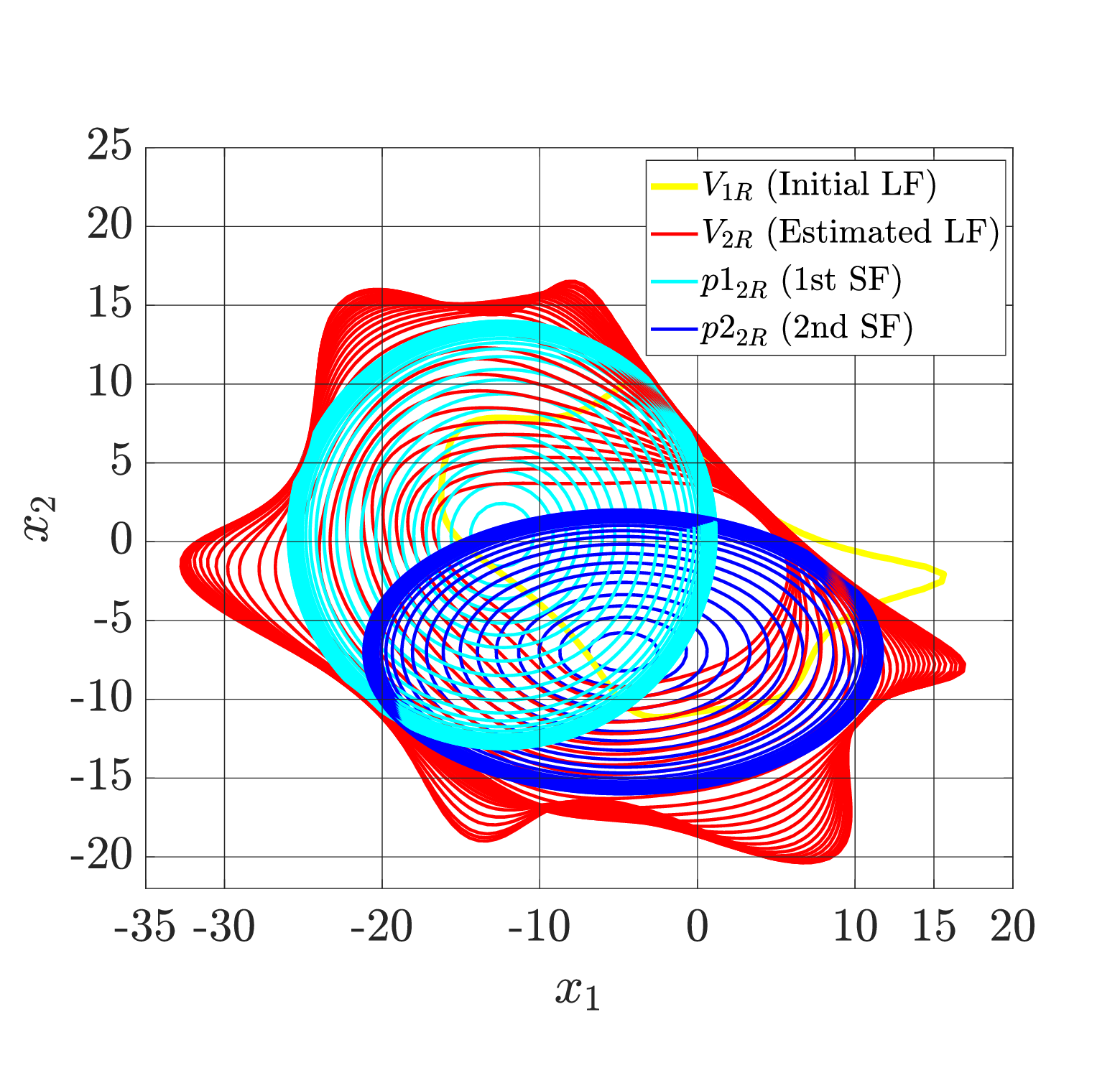}
		\label{7b_f}}	
	\subfigure[Comparison Among Estimated ROAs from all Rounds]{\includegraphics[scale=0.2]{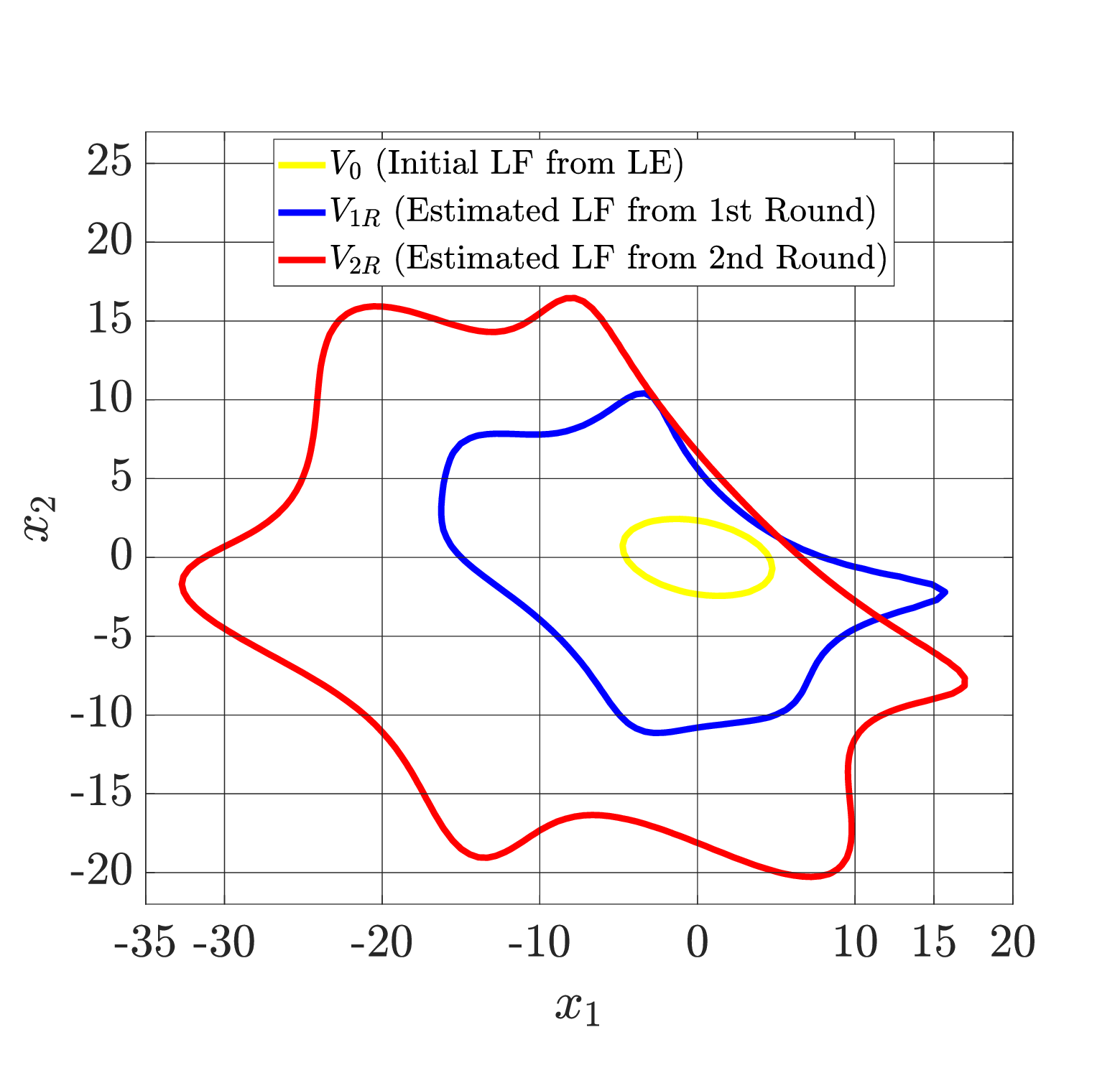}
		\label{7c_f}}\vfill
	\subfigure[Final Estimated ROA (Obtained from 2nd Round)]{\includegraphics[scale=0.2]{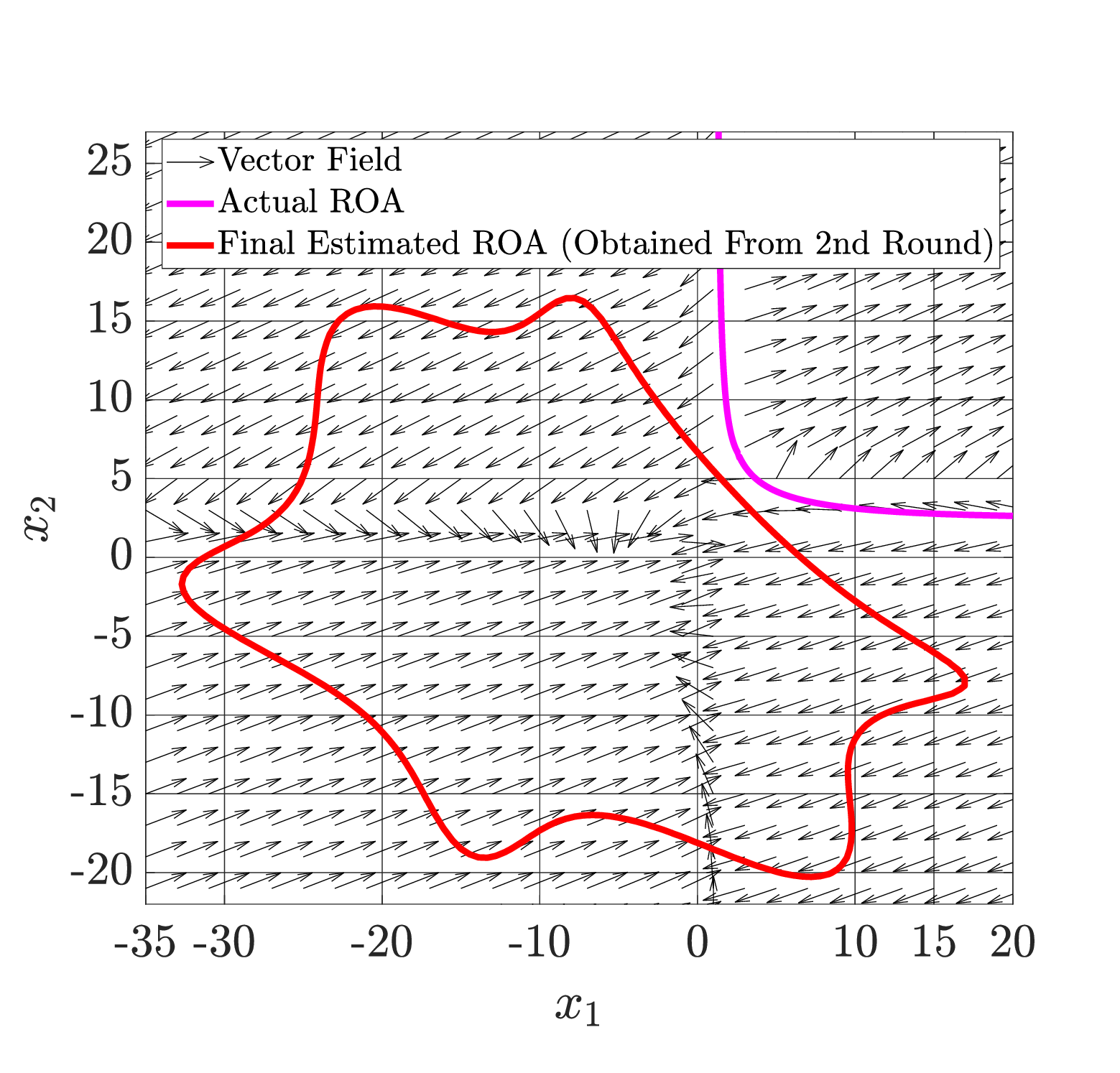}
		\label{7d_f}}
	\subfigure[Comparison Among Estimated ROAs based on Different Methods]{\includegraphics[scale=0.2]{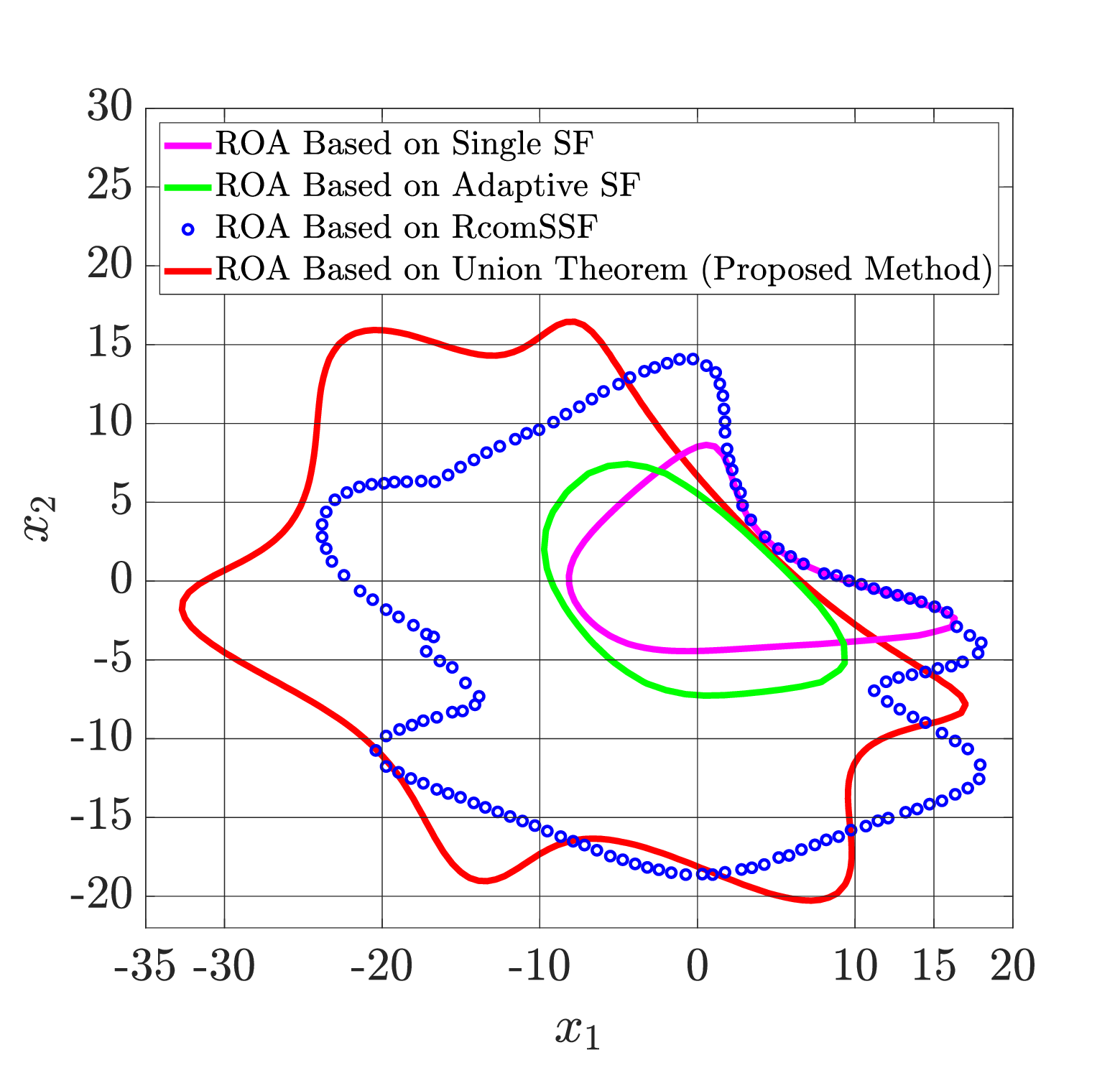}
		\label{7e_f}}
	\caption{ROA Estimation for Example 3}
	\label{7_f}\vfill	
	\subfigure[Estimated ROA from 1st Round]{\includegraphics[scale=0.2]{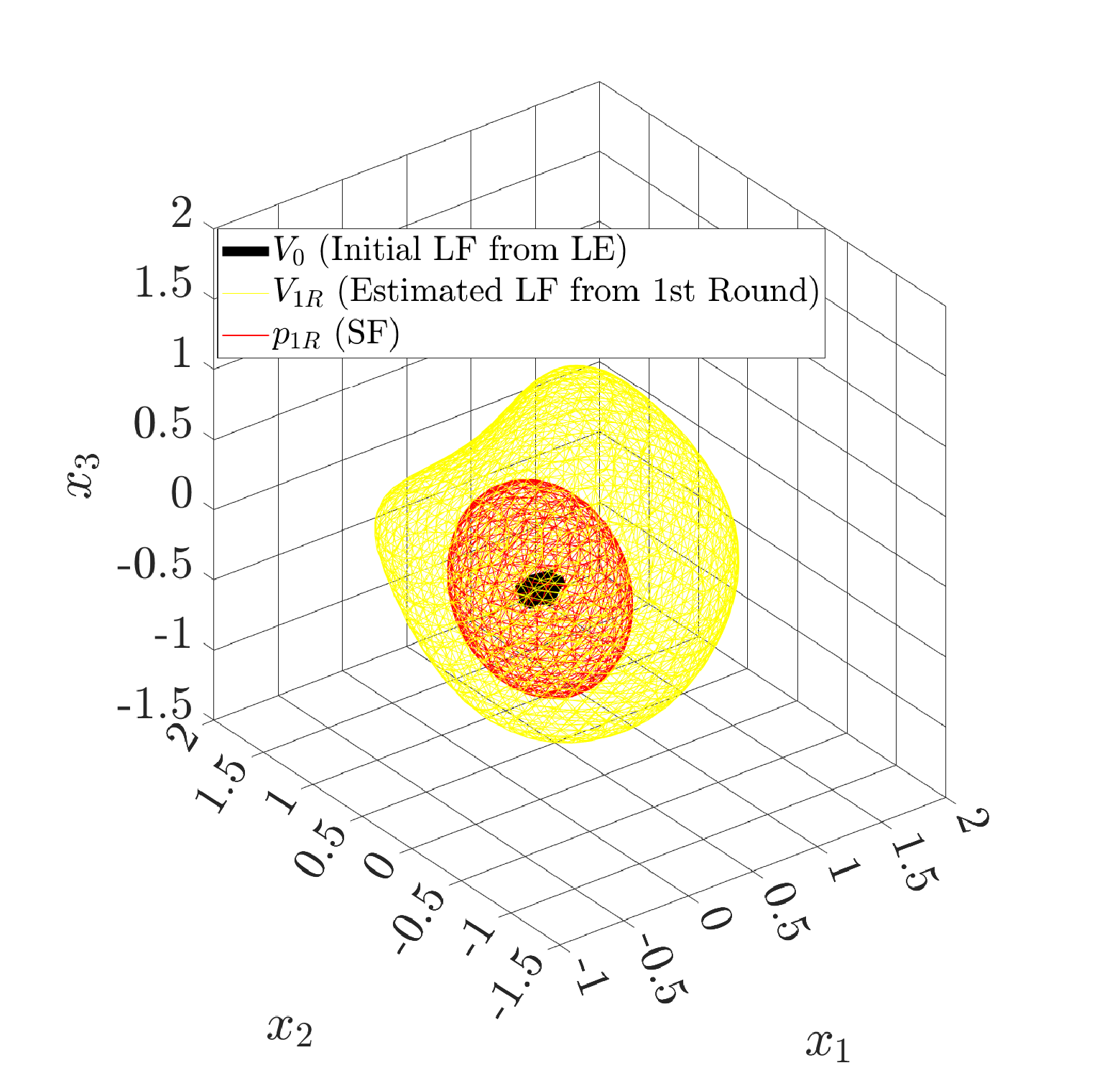}
		\label{8a_f}}
	\subfigure[Estimated ROA from 2nd Round]{\includegraphics[scale=0.2]{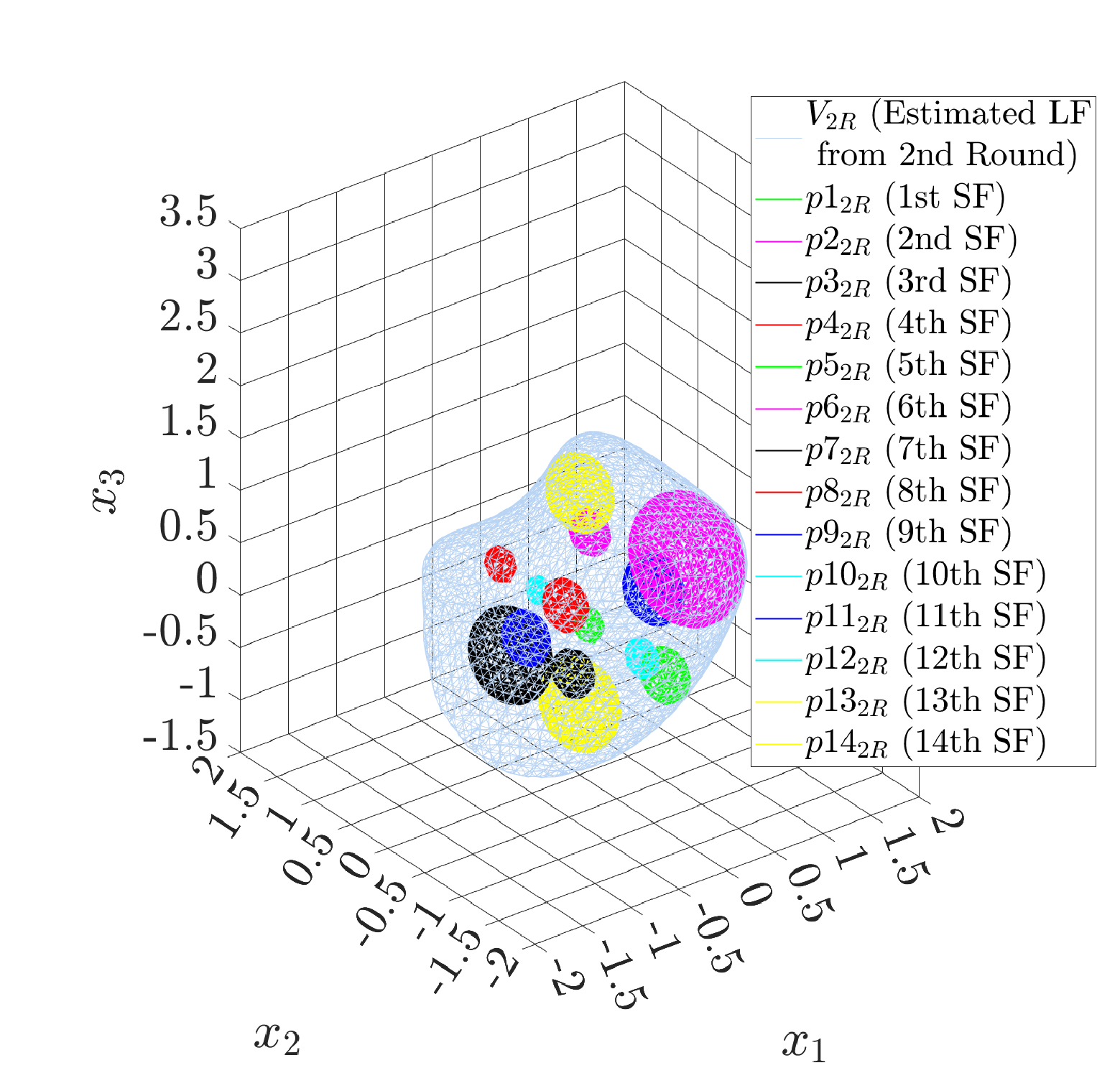}
		\label{8b_f}}
	\subfigure[Estimated ROA from 3rd Round]{\includegraphics[scale=0.2]{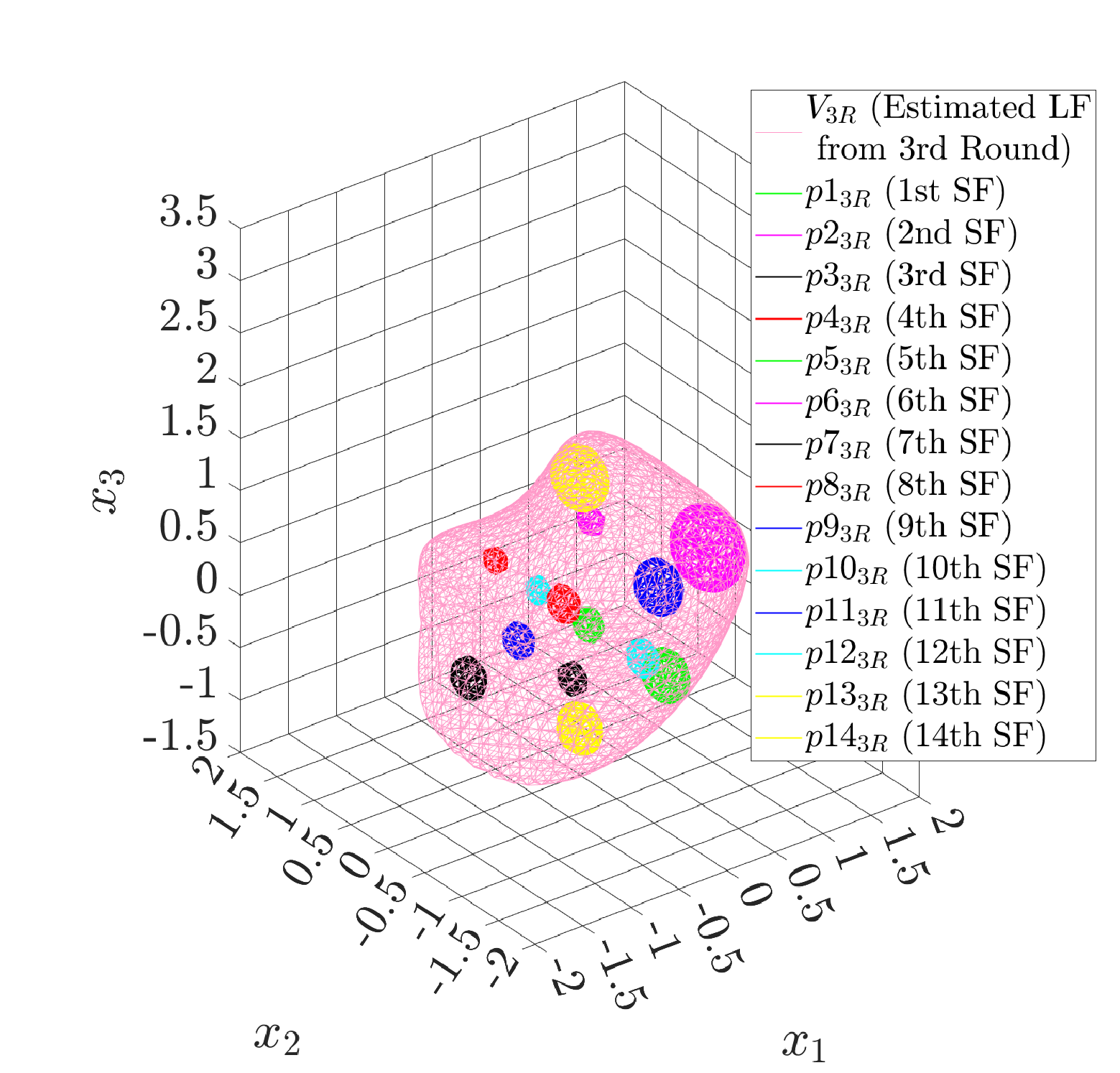}
		\label{8c_f}}\vfill
	\subfigure[Comparison Among Estimated ROAs from all Rounds]{\includegraphics[scale=0.2]{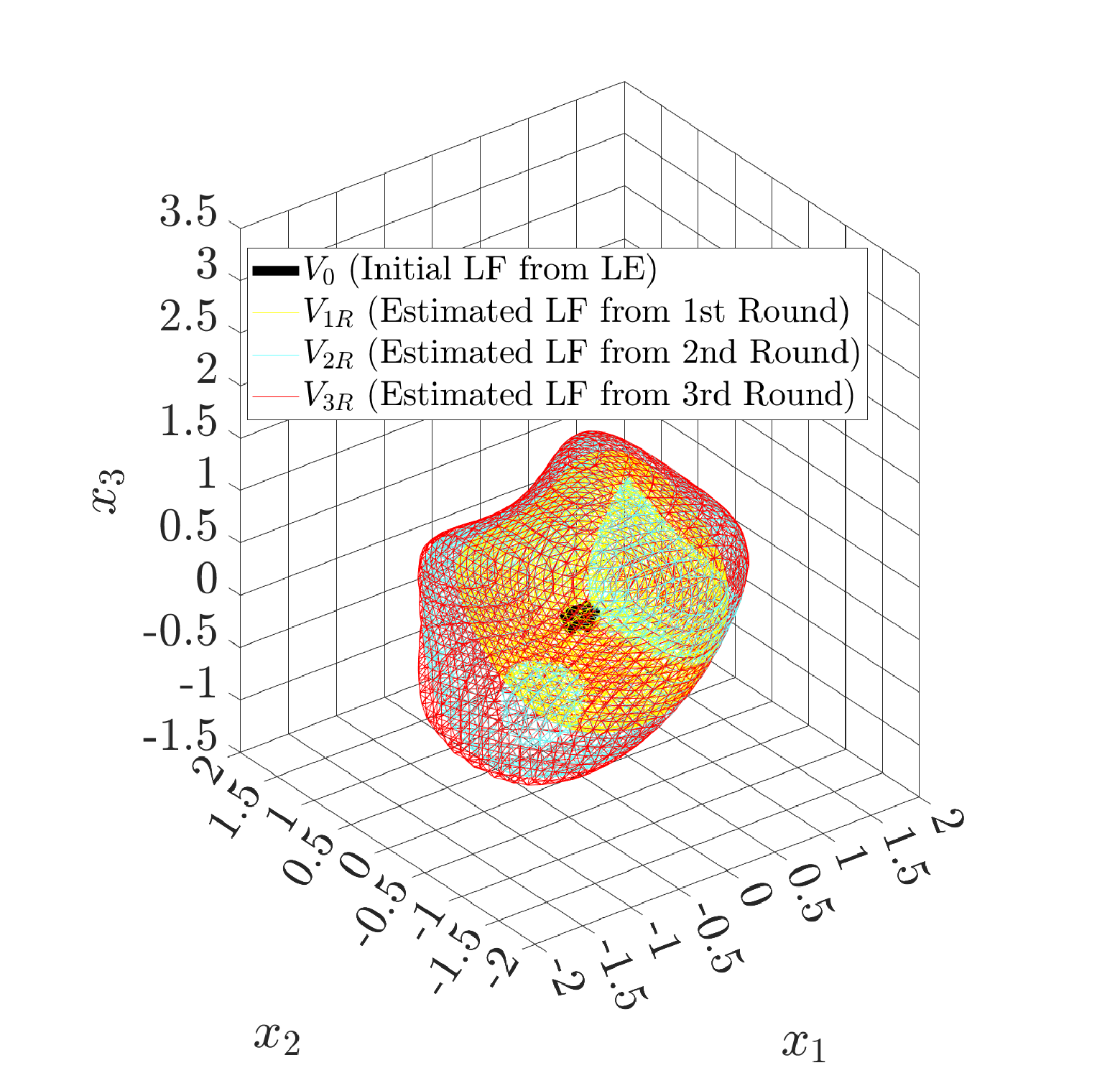}
		\label{8d_f}}
	\subfigure[Final Estimated ROA (Obtained from 3rd Round)]{\includegraphics[scale=0.2]{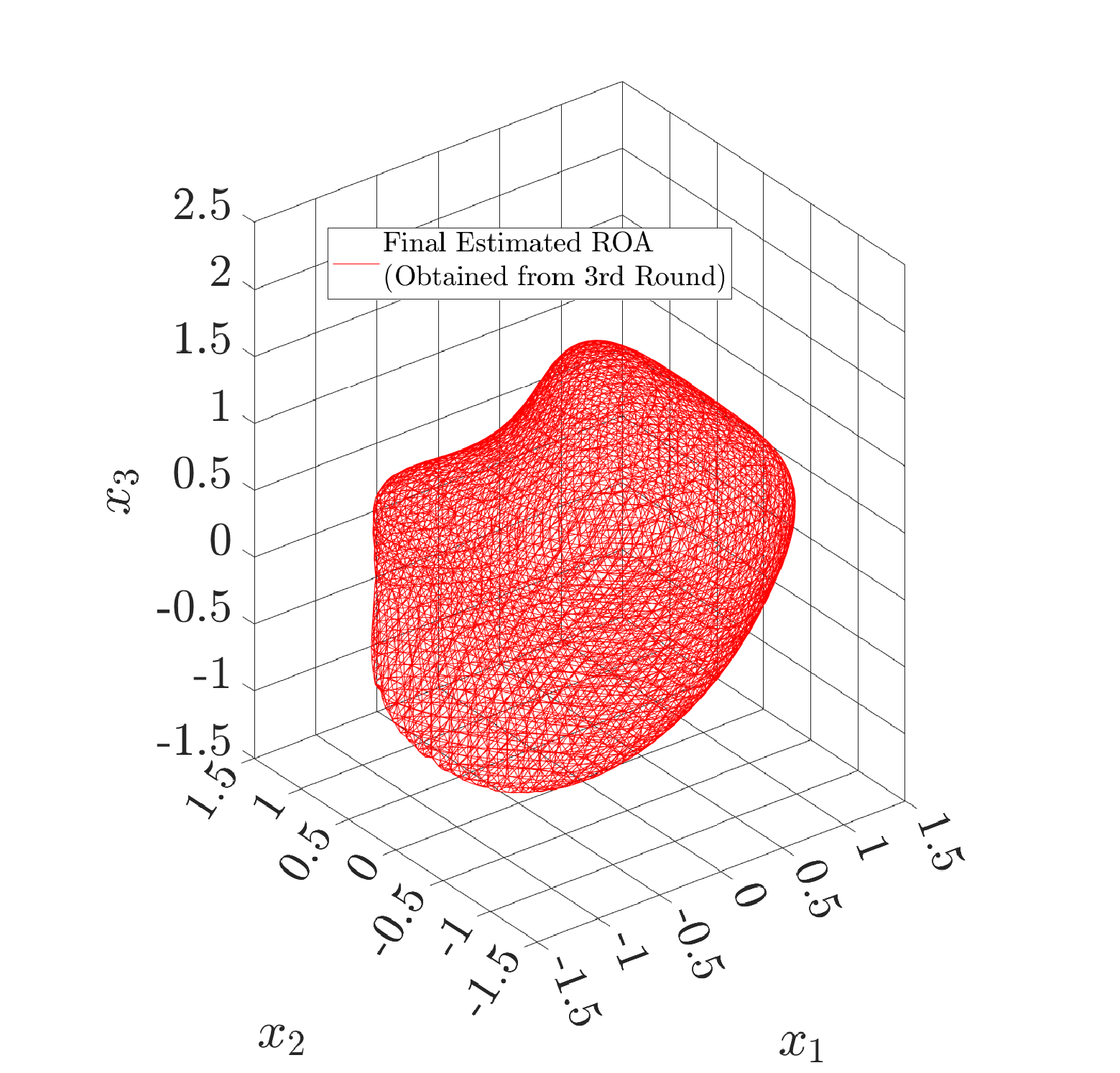}
		\label{8e_f}}
	\subfigure[Comparison Among Estimated ROAs based on Different Methods]{\includegraphics[scale=0.2]{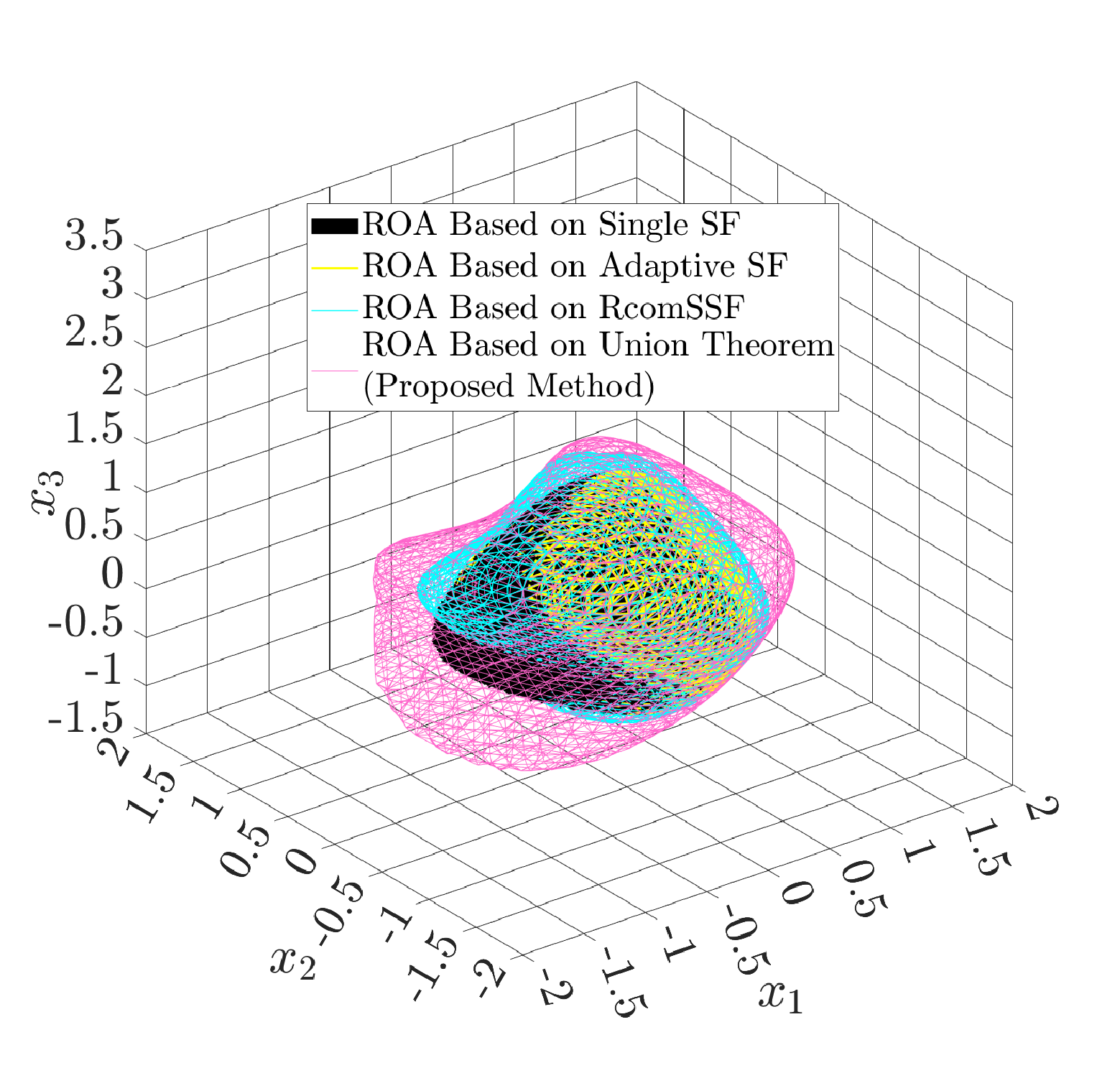}
		\label{8f_f}}
	\caption{ROA Estimation for Example 4}
	\label{8_f}
\end{figure*}

The proposed method's effectiveness is evaluated through extensive simulations using the SOSOPT Toolbox (\cite{r25} and \cite{r9}), on 2-D and 3-D systems with bounded or unbounded, symmetric, or non-symmetric ROAs. It should be noted that SeDuMi is required for the SOSOPT toolbox. An initial LF is necessary for each new round of iteration, obtained using the 'linstab' command of SOSOPT in the first round, which basically solve LE represented by Eq. \eqref{31b} and the estimated LF from the previous round in subsequent rounds. $l_1$ and $l_2$ are defined according to Eqs. \eqref{27a} and \eqref{25}, with $l_1=l_2=10^{-6}(x^Tx)$ for all examples.\\[3pt]
In each round of iteration, the center of the shifted shape function (given by equation \eqref{33}) must be determined, as it changes with each round. The number and type of shifted shape functions may vary, and different types may be used within the same round to improve the ROA estimation. The shape matrix $N$ is chosen using a combination of trial and error and insight from previous papers, \cite{r20} and \cite{r23}. The center $x^*$ is found by solving Eq. \eqref{35} for 2-D systems and Eq. \eqref{36} for 3-D systems, with predefined angles selected based on error and trial method. Additionally, $\sigma$ is set at $\sigma=0.8$ as stated in Eq. \eqref{35}. A six-degree LF [$(min(2), max(6))$] and fourth-degree $s_i$ [$deg(s_0)\rightarrow (min(2), max(4))$ and $deg(s_{1,2,...,n})\rightarrow (min(0), max(4))$] are used for all examples. The simulation stops when both stopping criteria A and B are met simultaneously. The result from the proposed method is compared to other methods, including an origin-centered SF, an adaptive SF, and RcomSSF.\\[3pt]
The proposed approach is advantageous in estimating a very large subset of the ROA with polynomial LF. However, The result depends on the type ($N$) and number of SF, and the selection of SF centers can impact the result. In this work, SFs and angles related to their center positioning are selected manually. A more structured approach for selecting these components would further improve the results.\\[3pt]
Results of methods represented by Eqs. \eqref{11} and \eqref{20} are shown prior to the proposed method's results to highlight their limitations and the need for the new method. These results are simulated using the Van der Pol system only due to its well-established stable boundary.\\[3pt]
\textbf{Example 1.} This first example is the well-known Van der Pol system, \cite{r23},
\begin{align}
	\begin{split}
		&\dot{x}_1=-x_2\\[2pt]
		&\dot{x}_2=x_1+5x_2(x_1^2-1)
	\end{split}
\end{align}
{\bf Results Based on Single and No SF:}
The results presented in this section are based on both the absence of SF and the use of a single SF. $N=[1\hspace{3pt}0;\hspace{1pt}0\hspace{3pt}0.5]$ is used for the method represented by Eq. \eqref{20}. In Fig.\ref{6_f0}, the actual ROA is shown by the black boundary, and the blue and green boundaries represent the ROA estimation based on no SF [Eq. \eqref{11}] and linearized dynamics (LD) [Eq. \eqref{31}], respectively. It can be noticed that no SF based method captured a very small region, even smaller than the ROA estimation based on LD. The red boundary represents the estimated ROA using a single SF [Eq. \eqref{20}], and the cyan color represents the corresponding SF. Among all the methods, the single SF-based method gives a better result. However, there still stay a significant untapped area between the estimated ROA based on single SF method, and the actual ROA. This untapped area can not be captured using the current single SF based method as the red and cyan curves almost touch the black curve in two points, further increments of the SF size will not give any new surrounded LF as it approaches the unstable region in these particular directions. Our proposed method presented in the next section addresses this issue.
{\bf Proposed Method outcome:} Here, we present the results using the proposed Union Theorem in the SOS optimization method. To simulate the results of this example, three rounds of iterations have been employed with three SFs used in each round, respectively. The parameters used for this process are given in Table \ref{T2}.
The results are presented in Fig. \ref{5_f}. Fig. \ref{5a_f}, Fig. \ref{5b_f}, and Fig. \ref{5c_f} display the LFs along with their corresponding SFs for each iteration. It is observable that a new LF is produced in every iteration when the sizes of the SFs are enlarged. It can be observed that the LF generated in the first round does not cover all areas, so it is necessary to go through additional rounds of iterations. In this case, three rounds of iterations have been used to capture the maximum area. A comparison of the results generated from all rounds is presented in Fig. \ref{5d_f}, and it can be noticed that the second and third rounds capture almost the same area. Furthermore, the results generated from the proposed method have been compared to simulated results from various other methods in Fig. \ref{5f_f}. One can see that the proposed method provides a much better approximation of the ROA than single SF and adaptive SF. The area predicted by the proposed method is comparable with RcomSSF, however, we need to remember that our method gives the polynomial final LF. The final result is presented in Fig. \ref{5e_f}, and it can be seen that the estimated ROA covers more than 99$\%$ of the actual region. The computational time for this simulation is 840 seconds, including 78, 46, and 5 iterations in the first, second, and third rounds of iteration, respectively.
\begin{table}[H]
	\caption{Simulation Parameters for Example 1\label{T2}}
	\centering
	\begin{tabular}{ | m{3em} | m{0.95cm}| m{1.4cm} | m{2.9cm} | }
		\hline
		Round &Initial LF$\hspace{2pt}(V_0)$& Angle\hspace{2pt}($\theta^{\circ}$) & Shape Matrix$\hspace{2pt}(N)$ \\ \hline
		\multirow{3}{*}{1st(1R)} &$V_0=$&$\theta1_{1R}\hspace{-2.5pt}=$60  & $N1_{1R}=[1\hspace{3pt}0;0\hspace{3pt} 0.5]$ \\\cline{3-4}
		&From& $\theta2_{1R}\hspace{-2.5pt}=$209 &  $N2_{1R}=[1\hspace{3pt}0;0\hspace{3pt} 0.5]$ \\\cline{3-4}
		&LE& $\theta3_{1R}\hspace{-2.5pt}=$260 &  $N3_{1R}=[1\hspace{3pt}0;0\hspace{3pt} 0.5]$ \\ \hline
		\multirow{3}{*}{2nd(2R)} &$V_0=$&$\theta1_{2R}\hspace{-2.5pt}=$60  & $N1_{2R}=[1\hspace{3pt}0;0\hspace{3pt} 0.5]$ \\\cline{3-4}
		&$V_{1R}$&$\theta2_{2R}\hspace{-2.5pt}=$209  & $N2_{2R}=[1\hspace{3pt}0;0\hspace{3pt} 0.5]$\\\cline{3-4}
		&& $\theta3_{2R}\hspace{-2.5pt}=$260  & $N3_{2R}=[1\hspace{3pt}0;0\hspace{3pt} 0.5]$ \\ \hline
		\multirow{3}{*}{3rd(3R)} &$V_0=$&$\theta1_{3R}\hspace{-2.5pt}=$60  & $N1_{3R}=[1\hspace{3pt}0;0\hspace{3pt} 0.5]$\\\cline{3-4}
		&$V_{2R}$& $\theta2_{3R}\hspace{-2.5pt}=$209  & $N2_{3R}=[1\hspace{3pt}0;0\hspace{3pt} 0.5]$ \\\cline{3-4}
		&& $\theta3_{3R}\hspace{-2.5pt}=$260  & $N3_{3R}=[1\hspace{3pt}0;0\hspace{3pt} 0.5]$ \\
		\hline
	\end{tabular}
\end{table}
\textbf{Example 2.} The following system \cite{r23} is considered for the method validation,
\begin{align}
	\begin{split}
		&\dot{x}_1=-4x_1^3+6x_1^2-2x_1,\\[2pt]
		&\dot{x}_2=-2x_2.
	\end{split}
\end{align}
The system has two stable sink nodes, $(0,0)$ and $(1,0)$, and a saddle point at $(0.5,0)$. There are two separate ROA corresponding to the two stable sink nodes. However, in this evaluation of the proposed method, only the equilibrium point $(0,0)$, whose ROA is unbounded in the $x_1<0.5$ plane, is considered. In order to simulate the outcome for this example, three rounds of iterations with three SFs in each round have been utilized and the following parameters of Table \ref{T3} are taken into account.
The results are shown in Fig. \ref{6_f}. Different types of SFs have been used for this example as it is bounded on one side and unbounded on the other. Therefore, the shape matrix is chosen so that it only grows in the unbounded directions. Figs. \ref{6a_f}, \ref{6b_f}, and \ref{6c_f} exhibit LFs with their corresponding SFs for each iteration, and a new LF is generated whenever SFs are increased. From Fig. \ref{6a_f}, it can be seen that the estimated region grows mainly in the $x_2$ direction, while it grows in the $-x_1$ direction in Fig. \ref{6b_f}. This is due to the different shape matrices used. In Fig. \ref{6c_f}, it can be seen that the region grows both in the $-x_1$ and $x_2$ directions due to the different shape matrices used within the same round of iteration. By following this process, it can be seen from Fig. \ref{6d_f} that the 3rd round of iteration captures a much larger region than the 1st and 2nd rounds of iteration. Fig. \ref{6e_f} shows the final estimated ROA obtained from the 3rd round of iteration, along with the vector field and boundary of the actual ROA. The total computational time for this simulation is 1230 seconds, including 32, 30, and 101 iterations in the 1st, 2nd, and 3rd rounds of iteration, respectively.  Fig. \ref{6f_f} clearly demonstrates that the proposed method yields a considerably larger ROA estimation while maintaining the polynomial LF. Although the estimation procedure was stopped, the process can be continued to obtain even larger ROA estimations.
\begin{table}[H]
	\caption{Simulation Parameters for Example 2\label{T3}}
	\centering
	\begin{tabular}{ | m{3em} | m{0.95cm}| m{1.4cm} | m{2.9cm} | }
		\hline
		Round &Initial LF$\hspace{2pt}(V_0)$& Angle\hspace{2pt}($\theta^{\circ}$) & Shape Matrix$\hspace{2pt}(N)$ \\ \hline
		\multirow{3}{*}{1st(1R)} &$V_0=$&$\theta1_{1R}\hspace{-2.5pt}=$132  & $N1_{1R}=[5\hspace{3pt}0;0\hspace{3pt} 0.3]$ \\\cline{3-4}
		&From& $\theta2_{1R}\hspace{-2.5pt}=$183 &  $N2_{1R}=[5\hspace{3pt}0;0\hspace{3pt} 0.3]$ \\\cline{3-4}
		&LE& $\theta3_{1R}\hspace{-2.5pt}=$234 &  $N3_{1R}=[5\hspace{3pt}0;0\hspace{3pt} 0.3]$ \\ \hline
		\multirow{3}{*}{2nd(2R)} &$V_0=$&$\theta1_{2R}\hspace{-2.5pt}=$132  & $N1_{2R}=[0.5\hspace{3pt}0;0\hspace{3pt} 1]$ \\\cline{3-4}
		&$V_{1R}$&$\theta2_{2R}\hspace{-2.5pt}=$183  & $N2_{2R}=[0.5\hspace{3pt}0;0\hspace{3pt} 1]$\\\cline{3-4}
		&& $\theta3_{2R}\hspace{-2.5pt}=$234  & $N3_{2R}=[0.5\hspace{3pt}0;0\hspace{3pt} 1]$ \\ \hline
		\multirow{3}{*}{3rd(3R)} &$V_0=$&$\theta1_{3R}\hspace{-2.5pt}=$132  & $N1_{3R}=[0.5\hspace{3pt}0;0\hspace{3pt} 1]$\\\cline{3-4}
		&$V_{2R}$& $\theta2_{3R}\hspace{-2.5pt}=$183  & $N2_{3R}=[5\hspace{3pt}0;0\hspace{3pt} 0.3]$ \\\cline{3-4}
		&& $\theta3_{3R}\hspace{-2.5pt}=$234  & $N3_{3R}=[0.5\hspace{3pt}0;0\hspace{3pt} 1]$ \\
		\hline
	\end{tabular}
\end{table}
\textbf{Example 3.} We now consider another second-order coupled nonlinear system given by the following equations \cite{r23}:
\begin{align}
	\begin{split}
		&\dot{x}_1=-50x_1-16x_2+13.8x_1x_2,\\[2pt]
		&\dot{x}_2=13x_1-9x_2+5.5x_1x_2.
	\end{split}
\end{align}
This example has a stable node at $(0,0)$ and a saddle point at $(1.45, 18.17)$. The unbounded ROA corresponding to the point $(0,0)$ has been studied utilizing the specific parameters of Table \ref{T4}.  To simulate this example's outcome, two rounds of iteration are considered, with three and two SFs used in the first and second rounds, respectively.
\begin{table}[H]
	\caption{Simulation Parameters for Example 3\label{T4}}
	\centering
	\begin{tabular}{ | m{3em} | m{0.95cm}| m{1.4cm} | m{3.55cm} | }
		\hline
		Round &Initial LF$\hspace{2pt}(V_0)$& Angle\hspace{2pt}($\theta^{\circ}$) & Shape Matrix$\hspace{2pt}(N)$ \\ \hline
		\multirow{3}{*}{1st(1R)} &$V_0=$&$\theta1_{1R}\hspace{-2.5pt}=$183  & $N1_{1R}=[15\hspace{3pt}0;0\hspace{3pt} 0.3]$ \\\cline{3-4}
		&From& $\theta2_{1R}\hspace{-2.5pt}=$183 &  $N2_{1R}=$ \\
		&LE&& $[14.47\hspace{3pt} 18.55;18.55\hspace{3pt} 26.53]$\\\cline{3-4}
		&& $\theta3_{1R}\hspace{-2.5pt}=$285 &  $N3_{1R}=[0.5\hspace{3pt}0;0\hspace{3pt} 12]$ \\ \hline
		\multirow{2}{*}{2nd(2R)} &$V_0=$&$\theta1_{2R}\hspace{-2.5pt}=$178  & $N1_{2R}=[1\hspace{3pt}0;0\hspace{3pt} 1]$ \\\cline{3-4}
		&$V_{1R}$& $\theta2_{2R}\hspace{-2.5pt}=$236  & $N3_{2R}=[0.3\hspace{3pt}0;0\hspace{3pt} 1]$ \\ 
		\hline
	\end{tabular}
 \end{table}
The results for this example are shown in Fig. \ref{7_f}. Figs. \ref{7a_f} and \ref{7b_f} show LFs and their corresponding SFs for each iteration, with each iteration generating a new LF as the SFs sizes grow.  A comparison among the ROAs obtained from all the rounds is shown in Fig. \ref{7c_f} and it can be noticed that the 2nd round of iteration captures a much larger region than the 1st round of iteration. Fig.~\ref{7d_f} shows the final estimated ROA obtained from the final round of iteration, along with the vector field and boundary of the actual ROA. The total computational time for this simulation is 174 seconds, including 20 and 31 iterations in the first and second rounds, respectively. Finally, the method is compared with other methods in Fig.~\ref{7e_f}. One can see that the proposed method overperforms the other considered techniques.\\[3pt]
\textbf{Example 4.} The following third-order system is  considered \cite{r23}:
\begin{align}
	\begin{split}
		&\dot{x}_1=x_2+x_3^2,\\[2pt]
		&\dot{x}_2=x_3-x_1^2-x_1(x_1-\frac{1}{6}x_1^3),\\[2pt]
		&\dot{x}_3=-x_1-2x_2-x_3+x_2^3+\frac{1}{10}(\frac{2}{3}x_3^3+\frac{2}{5}x_3^5).
	\end{split}
\end{align}
The system has an asymptotically stable equilibrium point at the origin $(0,0,0)$ and three other unstable equilibrium points. 3-D ellipsoidal SF is used for this system. The center of the shifted shape function was calculated using the 3-D linear equation given in Eq. \eqref{36}. As seen in Eq. \eqref{36}, two angles, the azimuth angle $\psi$ and the elevation angle $\theta$, are required to calculate the centers. To obtain these angles, a unit cuboid is considered, with its center placed at the equilibrium point $(0,0,0)$ of the system. Next, a line was drawn from the center of the cuboid to the vertex and middle point of each surface of the cuboid. The azimuth and elevation angles of these vertexes and middle points of the surfaces are then calculated using the known edge lengths (1 unit in all directions) of the cuboid. The cuboid has eight vertexes and six surfaces, resulting in a total of $14$ angles. The ellipsoidal shifted shape function and the angles (in degrees) are given below,
\begin{align}
	&p=\begin{bmatrix}
		x_1-x_1^*\\
		x_2-x_2^*\\
		x_3-x_3^*
	\end{bmatrix}
	N
	\begin{bmatrix}
		x_1-x_1^*\\
		x_2-x_2^*\\
		x_3-x_3^*
	\end{bmatrix};\hspace{5pt}
	N=\begin{bmatrix}
		1 & 0 & 0\\
		0 & 0.5 & 0\\
		0 & 0 & 0.5
	\end{bmatrix}\label{41}
\end{align}
	\begin{align}
		\begin{bmatrix}
			\psi\\
			\theta
		\end{bmatrix} &=
		\left[\begin{matrix}
			-45 & -45 &-135 & -135 & 45 & 45 & 135 & 135\\
			-35 & 35 & -35 & 35 & -35 & 35 & -35 & 35
		\end{matrix}\right.\nonumber\\
		&\qquad\qquad\qquad
		\left.\begin{matrix}
			0 & 90 & 180 &-90 & 0 & 0\\
			0 & 0 & 0 & 0 & 90 & -90
		\end{matrix}\right]\label{42}
	\end{align}
The primary eight angles represent the angles of the vertices, and the last six angles represent the angles of the middle points of the surfaces of the unit cuboid. To simulate the result, we have considered three rounds of iteration with one, and fourteen SFs used in the first, second, and third rounds, respectively. For this example, only the final result of each round is shown in the figures. In the first round of iteration, we used a single, origin-centered ellipsoid with a shape matrix equal to $N$ of Eq. \eqref{41}. For the remaining two rounds of iteration, we used the same angles of Eq. \eqref{42} and shape matrix of equation \eqref{41}. The results are shown in Fig.~\ref{8_f}. Fig.~\ref{8a_f}, Fig.~\ref{8b_f} and Fig.~\ref{8c_f} show the ROA estimation grows  from the 1st to the 3rd round. The results from all rounds are compared and shown in Fig.~\ref{8d_f}. The final result obtained from the third round is shown in Figure Fig.~\ref{8e_f}. The total computational time for this simulation is 2080 seconds, including 28, 19, and 5 iterations in the first, second, and third rounds of iteration, respectively. The obtained results are also compared with  results from  other methods in Fig.~\ref{8f_f}. Similar to the previous examples, using the proposed method one can obtain a better ROA approximation.
\section{Conclusion}
A  computationally-efficient numerical approach for estimating the Region of Attraction using the \emph{Union Theorem} is presented. The main novelty of this paper is the Union theorem that helps to generate a polynomial Lyapunov function produced by a set of shape functions. The main contribution of this paper is a numerical method for ROA estimations based on the Union theorem. A mathematical proof of the \emph{Union Theorem} along with its application to the numerical algorithm of ROA estimation is provided. The proposed method greatly improves ROA estimations and overcomes the limitations of the existing methods by providing the resultant Lyapunov function in a polynomial formulation. The proposed theorem and method facilitate further development of techniques for the design of control systems and construction of control Lyapunov function with enhanced ROA as well as certification using conventional SOS tools.  The method yields significantly enlarged ROA estimations even for systems with non-symmetric or unbounded ROA. The effectiveness of the proposed method is demonstrated through extensive simulation results for both 2-D and 3-D systems with bounded or unbounded, symmetric or non-symmetric ROAs. 

\section{Acknowledgement}
The first author acknowledges Government of India, Ministry of Social Justice and Empowerment, for supporting the studies under PhD scholarship no. 11015/33/2019 SCD-V NOS.

	\vspace{-50pt}
\begin{IEEEbiography}[{\includegraphics[width=1in,height=1.25in,clip,keepaspectratio]{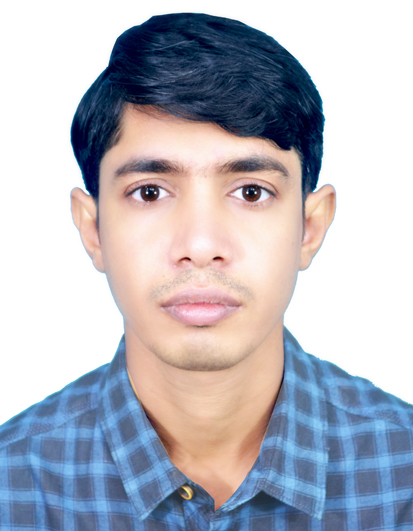}}]{Bhaskar Biswas} received the Bachelor of Engineering degree, in Aerospace Engineering, in 2014 from the Indian Institute of Engineering Science and Technology, Shibpur, and the Master of Technology, in Aerospace Engineering, in 2017 with specialization in dynamics and control from the Indian Institute of Technology, Bombay. He is currently pursuing a Ph.D. in Aerospace Engineering, with the Centre for Autonomous and Cyber-Physical Systems, at Cranfield University, UK. His research interests include guidance, stability analysis, machine learning, and control.
\end{IEEEbiography}
\begin{IEEEbiography}[{\includegraphics[width=1in,height=1.25in,clip,keepaspectratio]{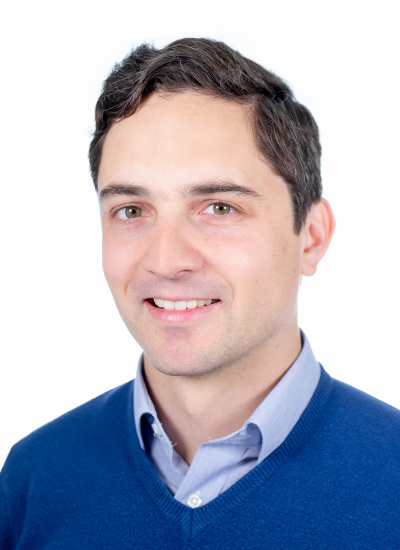}}]{Argyrios Zolotas} Dmitry Ignatyev received the BSc, MSc from Moscow Institute of Physics and Technology (MIPT), and the PhD degree in Applied Mathematics from MIPT in 2013. He is currently a Senior Research Fellow in Autonomous Systems and Control with the Centre for Autonomous and Cyber-Physical Systems, Cranfield University. His current research interests include nonlinear dynamics, stability analysis, machine learning and control.
\end{IEEEbiography}
\begin{IEEEbiography}[{\includegraphics[width=1in,height=1.25in,clip,keepaspectratio]{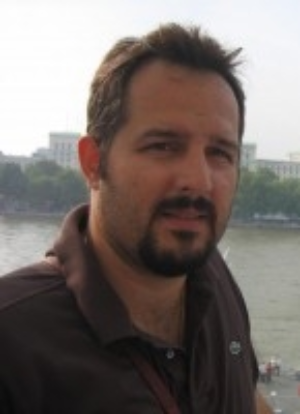}}]{Argyrios Zolotas} (S'96, M'04, SM'11) received the B.Eng.(Hons.) from the Univ. of Leeds, the M.Sc. degree from the Univ. of Leicester, the Ph.D. degree from Loughborough Univ. in the U.K. He is Reader in systems \& control at Cranfield University, heading the Autonomous Systems Dynamics \& Control research group within the Centre for Autonomous \& Cyber-Physical Systems. He actively participates in editorial duties for several journals. He has served as an Assoc. Editor for IEEE Trans. on Control Systems Technology, and for IEEE Control Systems Society Letters for several years.  He was Visiting Professor with GIPSA-Lab, Grenoble INP in 2018. He is a Fellow of the U.K. HEA, and a Fellow of the U.K. Institute of Measurement \& Control. His research focuses on advanced control, autonomous systems, and AI-based control applications.
\end{IEEEbiography}
\begin{IEEEbiography}[{\includegraphics[width=1in,height=1.25in,clip,keepaspectratio]{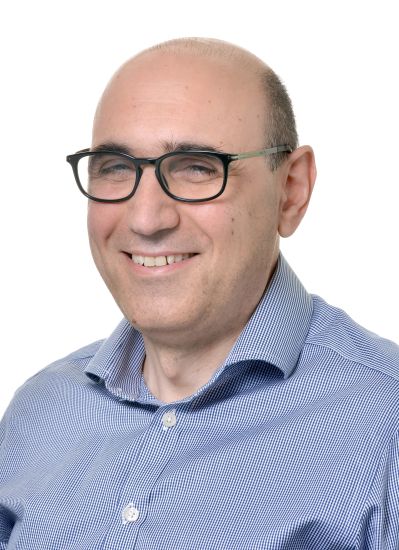}}]{Antonios Tsourdos} is a Professor of Autonomous Systems and Control with Cranfield University. He was appointed Head of the Autonomous Systems Group in 2007, Head of the Centre of Autonomous and Cyber-Physical Systems in 2012 and Director of Research - Aerospace, Transport and Manufacturing in 2015. Professor Tsourdos is chair of the IFAC Technical Committee on Aerospace Control, = member of the IMechE Mechatronics, Informatics And Control Group, the IET Aerospace Technical Network and the EPSRC KU-RAS Network. He is editorial board member for the Proceedings of the Institution of Mechanical Engineers, Part G: Journal of Aerospace Engineering, the Aerospace Science and Technology,  the Control Engineering Practice, the International Journal of Systems Science and the Journal of Intelligent and Robotic Systems. He has also served in the editorial board member of the IEEE Transactions on Aerospace and Electronic Systems and the IEEE Transactions of Instrumentation and Measurement.
\end{IEEEbiography}


\begin{thebibliography}{00}
\bibitem{r1}
Khalil, K. H.,
\newblock \emph{Nonlinear Systems}.
\newblock 3rd Edition, Prentice Hall, Inc., 2002.

\bibitem{r2}
Giesl, P.; Hafstein, S., 
\newblock ``Review on computational methods for Lyapunov functions,'' 
\newblock \emph{Discrete and Continuous Dynamical Systems - B}, Vol. 20, No. 8, pp. 2291-2331, 2015,
doi: 10.3934/dcdsb.2015.20.2291

\bibitem{r4}
VanAntwerp, J. G. and Braatz, R. D.,
\newblock ``A Tutorial on Linear and Bilinear Matrix Inequalities,''
\newblock \emph{Journal of Process Control}, Vol. 10, No. 4, pp. 363-385, 2000, https://doi.org/10.1016/S0959-1524(99)00056-6.

\bibitem{r5}
Tibken, B.,
\newblock ``Estimation of the Domain of Attraction for Polynomial Systems via LMIs,''
\newblock \emph{Proceedings of the 39th IEEE Conference on Decision and Control}, vol.4, pp. 3860-3864, 2000, doi: 10.1109/CDC.2000.912314.

\bibitem{r6}
Powers, V. and Wormann, T.,
\newblock ``An Algorithm for Sums of Squares of Real Polynomials,''
\newblock \emph{ Journal of Pure and Applied Algebra}, Vol. 127, No. 1, pp. 99-104, 2000, https://doi.org/10.1016/S0022-4049(97)83827-3.


\bibitem{r7}
Parrilo, P. A.,
\newblock \emph{Structured Semidefinite Programs and Semi-algebraic Geometry Methods in Robustness and Optimization,}
\newblock Ph.D. Thesis, California Institute of Technology, 2000, doi.10.7907/2K6Y-CH43.

\bibitem{r8}
Parrilo, P. A.,
\newblock ``Semidefinite Programming Relaxations for Semi-Algebraic Problems,''
\newblock \emph{Mathematical Programming}, 96(2):293–320, 2003, https://doi.org/10.1007/s10107-003-0387-5.

\bibitem{r9}
Balas, G.; Packard, A.; Seiler, P. and Topcu, U.,
\newblock \emph{Robustness Analysis of Nonlinear Systems}, 2009, \underline{http://users.cms.caltech.edu/\texttildelow utopcu/saveMaterial/LangleyWorkshop.html}
\bibitem{r10}
Peet, M. M.,
\newblock \emph{LMI Methods in Optimal and Robust Control}, 2020, \underline{http://control.asu.edu/MAE598\_frame.htm}.

\bibitem{r11}
Papachristodoulou, A. and Prajna, S., 
\newblock ``A Tutorial on Sum of Squares Techniques for Systems Analysis,''
\newblock \emph{Proceedings of the 2005, American Control Conference}, Vol. 4,  pp. 2686-2700, 2005, doi: 10.1109/ACC.2005.1470374.

\bibitem{r12}
Davis, J. M. and Eisenbarth, G.,
\newblock ``The Positivstellensatz and Nonexistence of Common Quadratic Lyapunov Functions,''
\newblock \emph{IEEE 43rd Southeastern Symposium on System Theory},
pp. 55-58, 2011, doi: 10.1109/SSST.2011.5753776.

\bibitem{r32}
Cunis, T.;  Legat, B.,
\newblock ``Sequential Sum-of-Squares Programming for Analysis of Nonlinear Systems,''
\newblock \emph{ArXivorg}, 2022, https://doi.org/10.48550/arXiv.2210.02142.


\bibitem{r25}
Seiler, P.,
\newblock ``SOSOPT: A Toolbox for Polynomial Optimization,''
\newblock 2016, https://doi.org/10.48550/arXiv.1308.1889.

\bibitem{r31}
Papachristodoulou, A.; Anderson, J.; Valmorbida, G.; Prajna, S.; Seiler, P.; Parrilo, P.; Peet, M. M. and Jagt, D.,
\newblock ``SOSTOOLS Version 4.00 Sum of Squares Optimization Toolbox for MATLAB,''
\newblock \emph{ArXivorg}, 2013, https://doi.org/10.48550/arXiv.1310.4716.

\bibitem{r13}
Ahmadi, A. A., 
\newblock \emph{Sum of Squares (SOS) Techniques : An Introduction}, 2016,
\underline{https://www.princeton.edu/{\raise.17ex\hbox{$\scriptstyle\sim$}}aaa/Public/Teaching}\discretionary{}{}{}\underline{/ORF523/ORF523\_Lec15.pdf}. 

\bibitem{r14}
Lall, S.,
\newblock \emph{Sums of Squares}, 2011, 
\underline{https://web.stanford.edu/class/ee364b}\discretionary{}{}{}\underline{/lectures/sos\_slides.pdf}.

\bibitem{r15}
Kunisky, D.,
\newblock \emph{Lecture Notes on Sum-of-Squares Optimization}, 2022, \underline{http://www.kunisky.com/static/teaching/2022spring-sos/sos-notes.pdf}.

\bibitem{r17}
Wloszek, Z. J.; Feeley, R.; Weehong, T.; Sun, K. and Packard, A.,
\newblock ``Control Applications of Sum of Squares Programming,''
\newblock \emph{Lecture Notes in Control and Information Sciences}, Vol. 312, pp. 580-508, 2005, https://doi.org/10.1007/10997703\_1.

\bibitem{r18}
Tan, W.; and Packard, A., 
\newblock ``Searching for Control Lyapunov Functions Using Sums of Squares Programming,''
\newblock \emph{42nd Annual Allerton Conference on Communications, Control and Computing}, pp. 210-219, 2004.
\bibitem{r19}
Jarvis-Wloszek, Z.,
\newblock \emph{ Lyapunov Based Analysis and Controller Synthesis for Polynomial Systems using Sum-of-Squares Optimization},
\newblock PhD thesis, University of California, Berkeley, 2003.

\bibitem{r20}
Tan, W.,
\newblock \emph{Nonlinear Control Analysis and Synthesis using Sum-of-Squares Programming},
\newblock PhD thesis, University of California, Berkeley, 2006.

\bibitem{r21}
Chakraborty, A.; Seilera, P. and Balasa, G.,
\newblock ``Nonlinear Region of Attraction Analysis for Flight Control Verification and Validation,''
\newblock \emph{Control Egineering Practice}, Vol. 19, No. 4, pp. 335-345, 2011, https://doi.org/10.1016/j.conengprac.2010.12.001.

\bibitem{r22}
Khrabrov, A.; Sidoryuk, M. and Ignatyev, D.,
\newblock ``Estimation of Regions of Attraction of Spin Modes,''
\newblock \emph{7th European Conference for Aerospace Sciences}, 2017, https://doi: 10.13009/EUCASS2017-312.

\bibitem{r34}
Li, D.; Tsourdos, A.; Wang, Z. and Ignatyev, D.,
\newblock ``Nonlinear Analysis for Wing-Rock System with Adaptive Control,'' \newblock \emph{Journal of Guidance, Control, and Dynamics}, Vol. 45, No. 11, pp. 2174-2181, 2022, https://doi.org/10.2514/1.G006775.

\bibitem{r23}
Li, D.; Ignatyev, D.; Tsourdos, A. and Wang, Z.,
\newblock ``Estimation of Non-Symmetric and Unbounded Region of Attraction using Shifted Shape Function and R-composition,'' 
\newblock \emph{ISA Transactions}, 2022, https://doi.org/10.1016/j.isatra.2022.11.015.

\bibitem{r24}
Balestrino, A.; Caiti, A.; Crisostomi, E. and Grammatico, S.,
\newblock ``R-composition of Lyapunov Functions,''
\newblock \emph{17th Mediterranean Conference on Control and Automation}, pp. 126-131, 2009, doi: 10.1109/MED.2009.5164527.


\bibitem{r26}
Sturm, J. F.,
\newblock ``Using SeDuMi 1.02, A Matlab Toolbox for Optimization Over Symmetric Cones,''
\newblock \emph{Optimization Methods and Software}, pp. 625-653, 1999, https://doi.org/10.1080/10556789908805766.

\bibitem{r27}
Sturm, J. F.,
\newblock \emph{SeDuMi Version 1.3}, 2001, \underline{https://sedumi.ie.lehigh.edu/?page\_id=58}.

\bibitem{r28}
Polik, I. and Terlaky, T.,
\newblock ``A Survey of the S-Lemma,''
\newblock \emph{SIAM Review}, Vol. 49, No. 3, pp. 371-418, 2007, http://www.jstor.org/stable/20453987.

\bibitem{r29}
Topcu, U. and Packard, A.,
\newblock ``Local Stability Analysis for Uncertain Nonlinear Systems,''
\newblock \emph{IEEE Transactions on Automatic Control}, Vol. 54, No. 5, pp. 1042-1047, 2009, doi: 10.1109/TAC.2009.2017157.

\bibitem{r30}
Khodadadi, L.; Samadi, B. and Khaloozadeh, H.,
\newblock ``Estimation of Region of Attraction for Polynomial Nonlinear Systems: A Numerical Method,''
\newblock \emph{ISA Transactions}, Vol. 53, No. 1, pp. 25-32, 2013, https://doi.org/10.1016/j.isatra.2013.08.005.



\bibitem{r33}
Vidyasagar, M.,
\newblock \emph{Nonlinear Systems Analysis},
\newblock 2nd Edition, Prentice Hall, Inc., 1993.
\end{thebibliography}
\end{document}